\documentclass[prl,aps, superscriptaddress,twocolumn]{revtex4-1}
\usepackage{amsmath, amsthm, amssymb,textcomp}
\usepackage{graphicx}
\usepackage{wasysym}
\usepackage{amsfonts}
\usepackage{bm}
\usepackage{enumerate}
\usepackage{color}
\usepackage[resetlabels]{multibib}
\usepackage{epstopdf}
\usepackage{latexsym}
\usepackage[breaklinks,colorlinks = true,linkcolor = red,urlcolor=cyan,citecolor=red]{hyperref}
\usepackage[caption=false,singlelinecheck=false]{subfig}
\DeclareMathOperator{\tr}{tr}
\usepackage[compat=1.1.0]{tikz-feynman}
\usepackage{times}
\usepackage{float}

\usepackage{times}
\newcommand{\bea}{\begin{eqnarray}}
\newcommand{\eea}{\end{eqnarray}}
\newcommand{\be}{\begin{eqnarray}}
\newcommand{\ee}{\end{eqnarray}}
\newcommand{\bw}{\begin{widetext}}
\newcommand{\ew}{\end{widetext}}

\def\ket#1{{|#1\rangle}}
\def\bra#1{{\langle #1 |}}

\begin{document}
\title{Topological d+s wave superconductors in a multi-orbital quadratic band touching system}
\author{GiBaik Sim}
\email{gbsim1992@kaist.ac.kr}
\affiliation{Department of Physics, Korea Advanced Institute of Science and Technology, Daejeon 305-701, Korea}
\author{Archana Mishra}
\email{amishra@kaist.ac.kr}
\affiliation{Department of Physics, Korea Advanced Institute of Science and Technology, Daejeon 305-701, Korea}
\author{Moon Jip Park}
\email{moonjippark@kaist.ac.kr}
\affiliation{Department of Physics, Korea Advanced Institute of Science and Technology, Daejeon 305-701, Korea}
\author{Yong Baek Kim}
\email{ybkim@physics.utoronto.ca}
\affiliation{Department of Physics, University of Toronto, Toronto, Ontario M5S 1A7, Canada}
\affiliation{Canadian Institute for Advanced Research, Toronto, Ontario, M5G 1Z8, Canada}
\affiliation{School of Physics, Korea Institute for Advanced Study, Seoul 02455, Korea}
\author{Gil Young Cho}
\email{gilyoungcho@postech.ac.kr}
\affiliation{Department of Physics, POSTECH, Pohang, Gyeongbuk 790-784, Korea}
\affiliation{School of Physics, Korea Institute for Advanced Study, Seoul 02455, Korea}
\author{SungBin Lee}
\email{sungbin@kaist.ac.kr}
\affiliation{Department of Physics, Korea Advanced Institute of Science and Technology, Daejeon 305-701, Korea}

\date{\today}
\begin{abstract}
Realization of topological superconductors is one of the most important goals in studies of topological phases in quantum materials. 
In this work, we theoretically propose a novel way to attain topological superconductors with non-trivial Fermi surfaces of Bogoliubov
quasiparticles. Considering the interacting Luttinger model with $j\!=\!3/2$ electrons, we investigate the dominant superconducting
channels for a multi-orbital quadratic band-touching system with finite chemical potential, which breaks the particle-hole
symmetry in the normal state. Notably, while the system generally favors d-wave pairing, the absence of the particle-hole symmetry necessarily
induces parasitic s-wave pairing. Based on the Landau theory with $SO(3)$ symmetry, we demonstrate that
two kinds of topological superconductors are energetically favored; uniaxial nematic phase with parasitic $s$ wave pairing ($d_{(3z^2-r^2)}\!+\!s$) and 
time-reversal-symmetry broken phase with parasitic $s$ wave pairing ($d_{(3z^2-r^2,xy)}\!+\!id_{x^2-y^2}\!+\!s$). 
These superconductors contain either nodal lines or Fermi pockets of gapless Bogoliubov quasiparticles and moreover exhibit topological winding numbers 
$\pm2$, leading to non-trivial surface states such as drumhead-like surface states or Fermi arcs. 
We discuss applications of our theory to relevant families of materials, especially half-heusler compound 
YPtBi, and suggest possible future experiments.
\end{abstract}
\maketitle


Recent surge of research activities on topological materials has paved novel avenues to obtain a variety of topological phases of quantum materials,
such as topological insulators, topological semimetals, and topological superconductors.\cite{fu2008superconducting,hasan2010colloquium,qi2011topological,wan2011topological,burkov2011topological,chiu2016classification}
One of the most pressing current issues is discovery and unambiguous confirmation of topological superconductors. While there exist a few
promising candidate materials, it would be great to identify generic material platforms, where different kinds of topological superconductors may
be obtained in a controlled fashion. In particular, topological superconductors with non-trivial gapless bulk excitations and exotic surface states
are of great interest. Such novel excitations hold the promise for future technological applications as they may appear at interfaces
between various topological superconductors and normal states.\cite{nayak2008non}

Superconductors with gapless Bogoliubov excitations may be obtained from Cooper pairs with non-zero angular momentum,
such as p-, d-, and f-wave superconductors.\cite{lee2008spin,mackenzie2003superconductivity,kuroki2001spin,sigrist1991phenomenological,sato2017topological} While these superconductors could be obtained via certain magnetic fluctuations in a single-band 
system, multi-orbital systems offer more generic routes to achieve such unconventional superconductors using inter-band pairing channels.\cite{moreo2009interband,boettcher2018unconventional}
Given the important roles of spin-orbit coupling in topological materials, various multi-orbital systems with heavy elements are proposed
for topological superconductors, which include $j\!=\!3/2$ system for half-heusler compounds and $j\!=\!5/2$ system for UPt$_3$.\cite{venderbos2017pairing,nomoto2016classification,yanase2016nonsymmorphic}

In this paper, we investigate emergent topological superconductors in the interacting Luttinger model, where $j=3/2$ multi-orbital electrons
form a quadratic band-touching at the Brillouin zone center. We consider a realistic situation of finite
chemical potential, which leads to breaking of the particle-hole symmetry in the normal state. 
We adopt the Landau theory of complex tensor order parameters including both $s$-wave and $d$-wave pairings, 
and study the Landau free energy functionals in terms of invariants under $SO(3)$ symmetry. 
Remarkably, the broken particle-hole symmetry has dramatic consequences for the nature of the superconducting states.
While the systems generally favors d-wave pairing channels, the broken particle-hole symmetry necessarily leads to
the presence of parasitic s-wave pairing. We obtain two leading topological superconducting phases; (i) uniaxial nematic phase with parasitic $s$ wave pairing ($d_{(3z^2-r^2)}\!+\!s$), where time-reversal, inversion and rotational symmetry along $z$ direction are preserved. (ii) time-reversal-symmetry broken phase with parasitic $s$ wave pairing ($d_{(3z^2-r^2,xy)}\!+\!id_{x^2-y^2}\!+\!s$), where only inversion and two-fold rotation along $z$ axis are preserved. 
These superconductors have gapless Bogoliubov quasiparticles and topological invariants (winding numbers $\pm2$) for each nodal line and Fermi pocket, respectively,
and also support non-trivial surface states. 

The selection of these superconducting states occur at the level of quartic interactions between order parameters.
This is distinct from the cases of pure $d$-wave paring states that may be chosen at the sixth order interactions between order parameters,
which was considered in previous studies of the Luttinger model with fine-tuned particle-hole symmetry or zero chemical potential.
Below we discuss details of our theoretical analyses and possible applications of our theory to superconductivity in relevant materials.
We also suggest possible ways to control the nature of the superconducting states using chemical doping, hydrostatic pressure, and temperature.     


We first introduce the multi-orbital Hamiltonian described by the Luttinger model with $j\!=\!3/2$ electrons.\cite{luttinger1955motion,savary2014new,moon2013non,boettcher2017anisotropy,yang2010topological}
The kinetic part of the Hamiltonian is given by 
\bea
h_0(\textbf{k})=\psi^\dagger_{\textbf{k}} \left [c_0k^2+\sum_{a=1}^5 c_a d_a(\boldsymbol{k})\gamma_a-\mu \right ] \psi_{\textbf{k}},
\label{eq:h}
\eea
where $\psi^\dagger_{\textbf{k}}\!=\!(\psi^\dagger_{\frac{3}{2},\textbf{k}},\psi^\dagger_{\frac{1}{2},\textbf{k}},\psi^\dagger_{-\frac{1}{2},\textbf{k}},\psi^\dagger_{-\frac{3}{2},\textbf{k}})$ is the four component spinor for $j\!=\!3/2$ electrons, $\mu$ is the chemical potential, $\gamma_{a}$ are the $4\times4$ gamma matrices and $d_a(\boldsymbol{k})\!=\!(\sqrt{3}/2)k_ik_jM^a_{ij}$ with  $3\times3$ real Gell-Mann matrices $M^a$. (See Section I of Supplementary Information (SI) for details)\cite{boettcher2017anisotropy}  
For the SO(3) symmetry, we set coefficients $c_a\!=\!c_1$. This model also has time-reversal and inversion symmetries. Especially when both $c_0\!=\!0$ and $\mu\!=\!0$, particle-hole symmetry is present in the system. Here and below, we consider $|c_0| \!<\! |\frac{c_1}{\sqrt{6}}|$ so that there is a single Fermi surface (with double degeneracy) for $\mu\!\neq\!0$. 
For interacting system, one can include additional terms,
\bea
h_{\text{int}}=g_0(\psi^\dagger\psi)^2 +  \sum_{a=1}^5  g_a (\psi^\dagger\gamma_a\psi)^2.
\label{eq:h_int}
\eea
The former corresponds to the onsite density-density interaction and the latter corresponds to the interactions between $d$-wave-orbital densities since $\psi^{\dagger}\gamma_a \psi$ transforms as $d$-wave orbitals or ``quadrupolar" moments (See Section I of SI).
Similar to the kinetic part of the Hamiltonian, we set coefficients $g_a\!=\!g_1$ for SO(3) symmetry. Note that these interactions correspond to so-called ``particle-hole" channel interactions and tend to condense the particle-hole composites $\sim \langle \psi^{\dagger} \psi \rangle$, when the coefficients are sufficiently negative. For instance, when $g_1\!<\!0$ becomes large, the system undergoes a nematic quantum phase transition toward a state with distorted Fermi surfaces. 

Here, we consider repulsive interactions, $g_0, g_a > 0$. We first use the Fierz identity for $j=3/2$ electrons to exactly decompose 
the particle-hole channel interactions into pairing channels. 
\begin{align}
h_{\text{int}} = g_s |\hat{\Delta}_s|^2  + g_d \sum_{a=1}^5 |\hat{\Delta}_{d, a}|^2, 
\end{align}
where $\gamma_{ab} \!\equiv\! i\gamma_{a}\gamma_{b}$, $\hat{\Delta}_s \!=\! \psi^T\gamma_{45}\psi$ is the s-wave pair that is invariant under all the symmetries of the system, and $\hat{\Delta}_{d, a} \!=\! \psi^T\gamma_{45}\gamma_a\psi, a = 1, 2, \cdots 5$ are the d-wave pairs that  transform exactly the same as the regular d-wave spin-singlet pair. From the Fierz identity, we find\cite{boettcher2018unconventional} 
\begin{align}
g_s=\frac{1}{4}(g_0+5g_1), ~~~ g_d=\frac{1}{4}(g_0-3g_1). 
\end{align} 
It is important to note that there is an instability toward superconducting states even when the bare interactions are all repulsive, i.e., $g_0\!>\!0$ and $g_1\!>\!0$, due to the minus sign relating $g_d$ and $g_1$. Hence, the system can have attractions only in the d-wave superconducting channels without the s-wave component. In the conventional single-band spin-1/2 electron system, the bare particle-hole channel interactions are neither decomposed exactly into the pairing channels nor favoring d-wave pairings.\cite{tinkham2004introduction}

Owing to such unique property, the Luttinger model has received lots of attention for the last couple of years for realzing unconventional d-wave superconducting states.\cite{agterberg2017bogoliubov,boettcher2018unconventional,roy2017topological,yu2018singlet} In fact, if the model is tuned to the particle-hole symmetric point with zero chemical potential, the system supports the pure $d$-wave superconducting state with nodal-line spectra beyond the critical value of the interactions.\cite{boettcher2018unconventional} However, we find that several drastic deviations from this expectation appear when this fine-tuned particle-hole symmetry is absent. To demonstrate this explicitly, below we consider the system with a finite chemical potential. We assume the attraction in the $d$-wave pairing $g_d$ is dominant compared to the $s$-wave pairing $g_s$, but keep the s-wave channel explicitly.
 
After integrating out the $j=3/2$ electrons, we now expand the Ginzburg-Landau free energy functional in terms of the superconducting order parameters $\Delta_s \!=\! \langle \hat{\Delta}_s \rangle$ and $\Delta_{a} \!=\! \langle \hat{\Delta}_{d,a} \rangle$. In the presence of the SO(3) symmetry, one defines symmetric traceless $3\!\times\!3$ tensor order parameter, $\phi \! \equiv \! \sum_{a}\Delta_a M^a$ with $M_a$ being the real Gell-Mann matrices. The free energy functional is written as,
\bea
\nonumber F&=&r_d|\vec{\Delta}|^2 +r_s|\Delta_s|^2 +q_{d_1}|\vec{\Delta}|^4 +q_{d_2}|\vec{\Delta}^2|^2 +q_{s}|\Delta_s|^4\\
\nonumber &+&m_2(|\vec{\Delta}|^2|\Delta_s|^2) +m_3(\vec{\Delta}^2(\Delta_s^*)^2+c.c.) \\
&+&q_{d_3}\text{tr}((\phi^\dagger\phi)^2) +m_1(\text{tr}(\phi^2\phi^\dagger)\Delta_s^*+c.c.).
\label{eq:freet}
\eea
Here, $\vec{\Delta}$ represents the vector $(\Delta_1, \Delta_2, \cdots, \Delta_5)$ for d-wave pairing. $(\Delta_1,\Delta_2)$ are d-wave pairings $(d_{x^2-y^2},d_{3z^2-r^2},)$ with $e_g$ symmetry and $(\Delta_3,\Delta_4,\Delta_5)$ are for d-wave parings $(d_{yz},d_{xz},d_{xy})$ with $t_{2g}$ symmetry, respectively.
A few remarks follow: We assume that there is a range of parameters such that $r_s\! >\!0$ and $r_d \!<\!0$ with dominant attractive $d$-wave channels. Within one-loop expansion, $q_{d_3}$ is always zero and so we drop it from here (See Section II of SI). The last $m_1$ term which couples d-wave and s-wave pairs, is zero within the leading one-loop computation in the presence of particle-hole symmetry, i.e. $m_1\!=\!0$ for $\mu\!=\! c_0 \!=\!0$ (See Section II of SI for details). The absence of $m_1$ term allows the previous studies to access the pure d-wave superconductors. However, when the fine-tuned particle-hole symmetry is absent so that $m_1$ is finite, we find that the non-zero d-wave condensates $\vec{\Delta} \!\neq\! 0$ always induce the parasitic s-wave superconductivity ($\Delta_s \!\neq\!0$) and change the nature of the superconducting states. 

\begin{figure}[t]
 	\includegraphics[width=1\columnwidth]{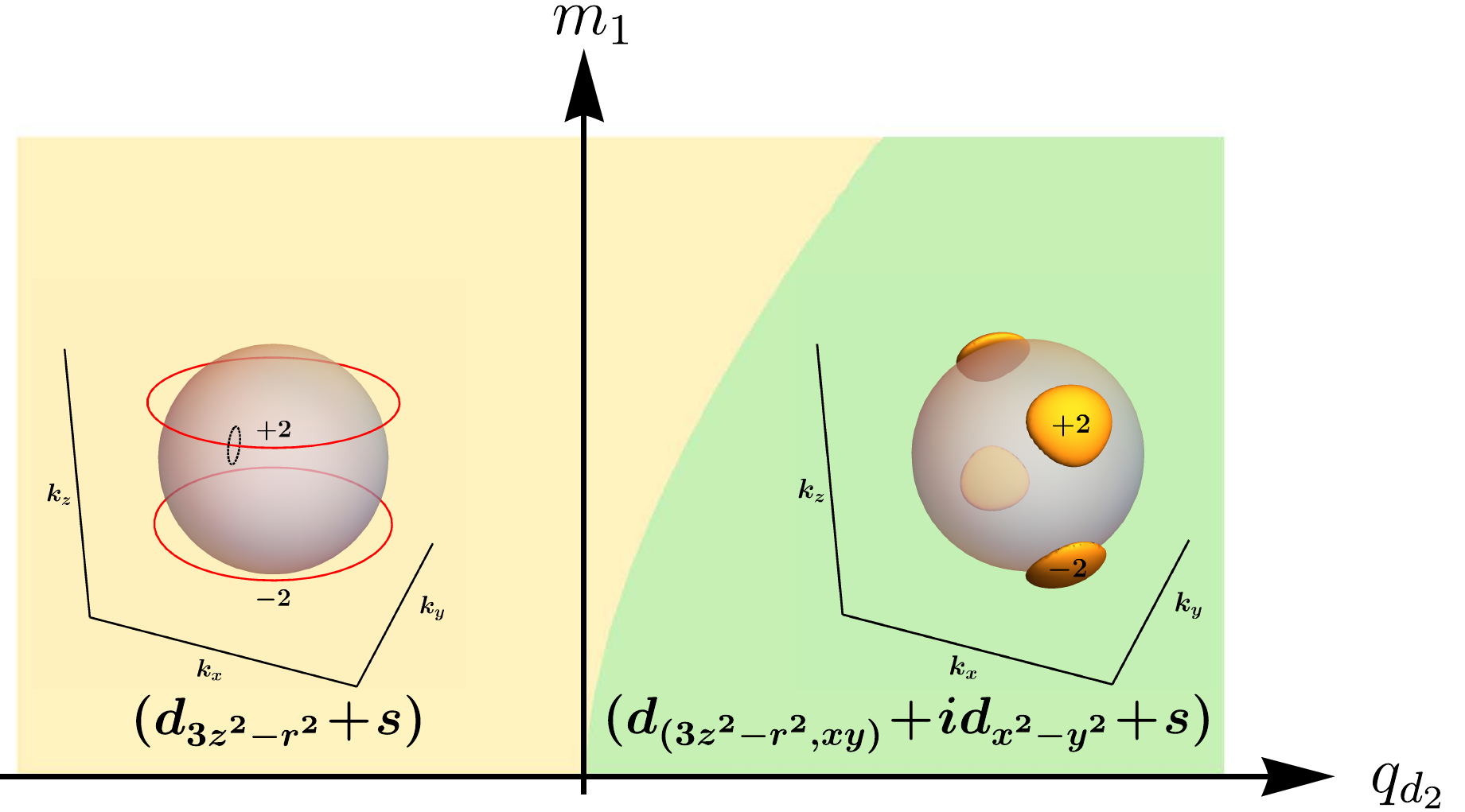}
 	\caption{(color online) Phase diagram of topological superconductors with $d$-wave and parasitic $s$-wave pairs, as a function of $q_{d_2}$ and $m_1$ in Eq.\eqref{eq:freet}. Interplay between $q_{d_2}$ term and $m_1$ term drives the system into topologically distinct phases: time reversal symmetric phase with $(d_{3z^2-r^2}\!+\!s)$ pairing (yellow) and time reversal symmetry broken phase with $(d_{(3z^2-r^2,xy)}\!+\!id_{x^2-y^2}\!+\!s)$ pairing (green). Each inset shows the gap structure of each phase, with topological invariant of each nodal ring or Bogoliubov Fermi pocket. Semi-transparent sphere indicates the normal-state Fermi surface. See main text for more details.
 	}
 	\label{fig:pdwm}
 \end{figure}

With these in mind, we look for the global phase diagram by minimizing the free energy in Eq.\eqref{eq:freet}. In principle, one should compute the free energy for all possible configurations of six distinct complex fields. However, we can use the SO(3) rotation symmetry to reduce the number of these symmetry-breaking patterns down to only a few configurations. In particular, if time-reversal symmetry is present, all the pairing fields become real. In such case, all the distinct symmetry-broken states can be obtained from the rotations of the three fields $(\Delta_1, \Delta_2, \Delta_s )$.\cite{boettcher2018unconventional} Similarly, in the absence of time-reversal symmetry, one can construct all the distinct symmetry-broken phases by the two complex fields $(\Delta_1,\Delta_2)\! \in \! \mathbb{C}^2$ and four real fields $(\Delta_3,\Delta_4,\Delta_5,\Delta_s) \! \in \! \mathbb{R}^4$ (See Section III of SI for details).

We now numerically minimize the free energy. In the presence of time-reversal symmetry, we find that the lowest-energy state is given by $\vec{\Delta}\!=\!(0,\Delta_2,0,0,0)$ with parasitic $\Delta_s$, i.e., uniaxial nematic phase with subdominant s-wave $(d_{3z^2-r^2}\!+\!s)$. 
Within the real manifolds of $\vec{\Delta}$, one can also perform tensorial differentiation and show that the stable equilibrium phase is indeed uniquely determined to be the uniaxial nematic state with parasitic $s$-wave pairing\cite{de2008landau} (See Section III of SI for details). 
On the other hand, without time-reversal symmetry, three components of the d-wave pairings contribute, which are parametrized as $(i\Delta_1,\Delta_2,0,0,\Delta_5)$ with real values of $\Delta_1,\Delta_2$, and $\Delta_5$. Again these d-wave pairings induce parasitic s-wave pairing and so we finally find the ($d_{(3z^2-r^2,xy)}\!+\!id_{x^2-y^2}\!+\!s$) state as the ground state. This state respects two fold rotation $C_{2z}$ and inversion $\mathcal{P}$ symmetries.

Without $m_1$ term in Eq.\eqref{eq:freet}, the $\pm$ sign of $q_{d_2}$ simply favors either complex $d\!+\!id$ or real $d$ wave pair respectively. 
However, the presence of $m_1$ term leads to particular choices of $d$ wave pairing with parasitic $s$ wave, i.e., $(d_{3z^2-r^2}\!+\!s)$ or $(d_{(3z^2-r^2,xy)}\!+\!id_{x^2-y^2}\!+\!s)$ as mentioned above. 
Such $m_1$ term becomes finite with particle-hole symmetry breaking and its magnitude is proportional to the chemical potential $\mu$ and $c_0$ (See Section II of SI for details).  Furthermore, the $m_1$ term induces competition between $|\vec{\Delta}^2|^2$ and $(\text{tr}(\phi^2\phi^\dagger)\Delta_s^*+c.c.)$, which favors uniaxial nematic phase with $(d_{3z^2-r^2})$ pairing for any sign of $m_1$. This makes broader region in the phase diagram for the time reversal symmetric phase with $(d_{3z^2-r^2}\!+\!s)$ pairing, in comparison to the time-reversal-symmetry broken phase with $(d_{(3z^2-r^2,xy)}\!+\!id_{x^2-y^2}\!+\!s)$ pairing. Fig.\ref{fig:pdwm} shows the phase diagram as a function of $m_1$ and $q_{d_2}$ while keeping $q_{d_1}$, $q_s$, $m_2$ and $m_3$ constant. Indeed, the region for uniaxial nematic phase with $(d_{(3z^2-r^2)}\!+\!s)$ pair gets wider with larger value of $m_1$. Similar uniaxial nematic phase has also been discussed in Ref.\onlinecite{boettcher2018unconventional}, where this phase is stabilized at the sixth order level within pure $d$ wave pairing order parameters. We emphasize that, in our study, the parasitic $s$ wave pairing is crucial for the special selection of $d$ wave, the uniaxial nematic phase, at quartic order via $(\text{tr}(\phi^2\phi^\dagger)\Delta_s^*+c.c.)$. 


In terms of gap structure, the multi-orbital nature of the superconducting states has striking consequences, namely the presence of gapless Bogoliubov quasiparticles with nodal lines
and Fermi surfaces.\cite{brydon2018bogoliubov,bzduvsek2017robust} In conventional single-band case, the parasitic s-wave pairing will fully gap out the Fermi surface and leave no gapless excitation for any dominant d-wave pairing configurations. In the $j\!=\!3/2$ multi-orbital model, however, we will see that the fermions remain gapless even in the presence of s-wave pairing. In particular, for the time-reversal symmetric state, the gapless nodal line is present, while for the time-reversal broken state, the gapless fermions form Fermi surfaces. Remarkably, these gapless Bogoliubov quasiparticles have topological character with a nonzero winding number or Chern number. This results in unique surface states, either drumhead states or Fermi arcs at the surface boundary, which we discuss below.\cite{bzduvsek2017robust}

\begin{figure}[b]
 	\includegraphics[width=1\columnwidth]{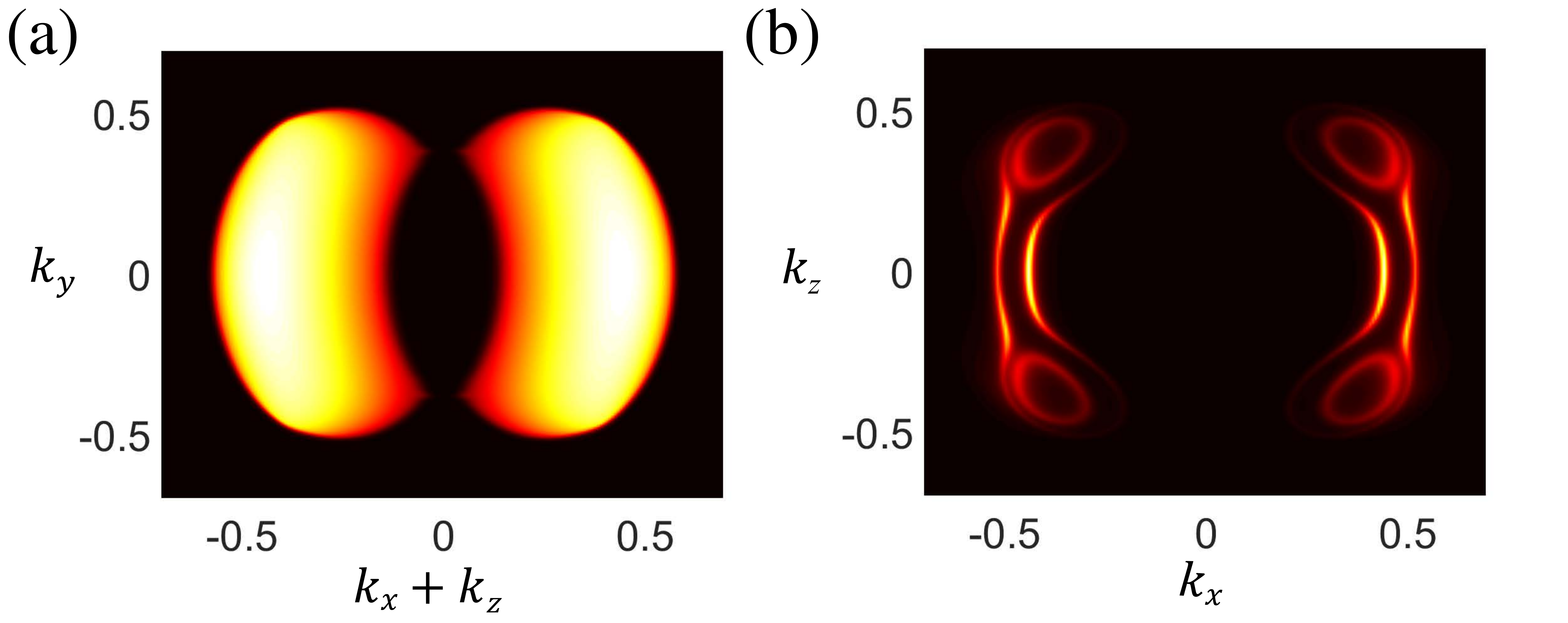}
 	\caption{(color online) Surface states of (a) $(d_{3z^2-r^2}\!+\!s)$ superconductor with topological line nodes having winding number $\pm2$; At $(10\bar{1})$ surface, 
	drumhead-like surface states are shown in the region where the projections of both line nodes with opposite winding numbers do not overlap. (b) $(d_{(3z^2-r^2,xy)}\!+\!id_{x^2-y^2}\!+\!s)$ superconductor with topological Fermi surface having Chern numbers $\pm2$; Fermi arcs are shown at $(010)$ surface, which connect two bulk Fermi surfaces with opposite Chern numbers. See main text for more details.
 	}
 	\label{fig:surface}
 \end{figure}

To demonstrate this explicitly, we study the Bogoliubov-de Gennes (BdG) Hamiltonian $H \!=\! \sum_{\textbf{k}}\Psi_{\textbf{k}}^{\dagger}\mathcal{H}(\textbf{k})\Psi_{\textbf{k}}$ with the Hamiltonian matrix $\mathcal{H}(\textbf{k})$ defined as,
\bea
\mathcal{H}(\textbf{k})=\left(
\begin{array}{cc}
	h_0(\textbf{k}) & \Delta \\
	\Delta^{\dagger} & -h_0^{T}(-\textbf{k}) \\
\end{array}\right).
\label{H}
\eea
Here, $\Psi_{\textbf{k}}\!=\!(\psi_{\textbf{k}}^T,\psi_{\textbf{-k}}^{\dagger})^T$ where $\psi_{\textbf{k}}$ is defined in Eq.\eqref{eq:h} and $\Delta\!=\!\gamma_{45}\Delta_s\!+\!\sum_a \gamma_a\gamma_{45}\Delta_a$. 
We find the momentum points where Bogoliubov quasiparticles become gapless, using Pfaffian $P({\textbf{k}})$ that satisfies the relation $P({\textbf{k}})^2 \!=\!\text{det}[{\mathcal{H}}(\textbf{k})]$ (See Section IV of SI for details.)\cite{brydon2016pairing,bzduvsek2017robust} 
For the time-reversal symmetric superconductor with $(d_{3z^2-r^2}\!+\!s)$ pairing, four-fold degenerate line nodes appearing at $k_z \!=\! \pm k_0$, where $k_0$ is a function of $(\Delta_s, \Delta_2)$ and Fermi momentum $k_F$. 
In Fig.\ref{fig:pdwm}, the left inset shows such line nodes. 
Furthermore, the winding numbers are $\pm2$ for each line-node located at $k_1\!=\!\pm k_0$, respectively, which is consistent with $2\mathbb{Z}$ classification of 
the nodal lines studied in Ref.\onlinecite{bzduvsek2017robust}. Of course, when s-wave pairing becomes large enough, the system is fully gapped (See Section V of SI for details). Fig.\ref{fig:surface} (a) shows the drumhead-like surface state for such topological line nodes of Bogoliubov quasiparticles. To observe this, we show the $(10\bar{1})$ surface so that there exist regions where the projections of the line nodes with opposite $\pm2$ winding numbers do not overlap. 

On the other hand, the time-reversal-symmetry broken superconductor with $(d_{(3z^2-r^2,xy)}\!+\!id_{x^2-y^2}\!+\!s)$ pairing has multiple Fermi pockets with two-fold degeneracy. 
Remarkably, each of these pockets has topological characteristic with an even Chern number. In this case, our system belongs to the class D, and according to the classification, each Fermi pocket is characterized by the two invariants $(l , n) \!\in\! \mathbb{Z}_2 \!\times \!2\mathbb{Z}$.\cite{bzduvsek2017robust,brydon2018bogoliubov} Here the integer index $n\!\in\! 2\mathbb{Z}$ corresponds to the Chern number, which characterizes the winding number around the Fermi pockets. As in the Weyl nodes, such winding numbers imply the presence of the surface Fermi arc states.\cite{wan2011topological,schnyder2015topological} Our Fermi pockets have the invariants $(1, \pm 2)$ and thus we have the zero-energy arc states connecting the pockets on the surface. The right inset of Fig.\ref{fig:pdwm} shows the Fermi pockets which preserve  $C_{2z}$ rotation and inversion $\mathcal{P}$. There are four distinct Fermi pockets; two of them located at $k_z\!>\!0$ have the Chern number +2 and other two at $k_z\!<\!0$ have the Chern number -2. Fig.\ref{fig:surface} (b) shows the Fermi arcs at $(010)$ surface which connect two bulk Fermi pockets having $\pm2$ Chern numbers. 
With increasing $s$ wave pairing, these Fermi pockets evolve and change their topology. They are eventually gapped out with sufficiently large $s$ wave. (See Section V of SI for details.)


To recap our discussion above, superconductivity in the Luttinger model prefers topologically non-trivial d-wave pairings with parasitic s-wave component;
the time-reversal-invariant uniaxial nematic phase with $(d_{3z^2-r^2}\!+\!s)$ pairing and the time-reversal-symmetry broken superconductor with $(d_{(3z^2-r^2,xy)}\!+\!id_{x^2-y^2}\!+\!s)$ pairing. As long as the Fermi level is not exactly at the quadratic band-touching point, the subdominant s-wave pairing always makes the system to select these particular choices of 
d-wave pairings. 
It is important to note that the nature of such superconducting states can easily be changed by tuning the chemical doping or applying hydrostatic pressure, with the underlying assumption of small Fermi surface, i.e, small $k_F$. In particular, chemical doping is directly related to the magnitude of $m_1$ coefficient in the Landau free energy functional,
which can be used to drive the transition between two topological superconductors, as shown in the phase diagram of Fig.\ref{fig:pdwm}. Furthermore, applying hydrostatic pressure may change the sign of mass term $r_s$ in Eq.\ref{eq:freet} so that the ratio of pairing order parameters between $d$ wave and $s$ wave can be controlled, which will eventually change  the nature of the superconducting states and their topological characteristics. The evolution of the phase diagram with temperature would be another avenue to explore possible interplay between different topological superconducting states. (See Section V of SI for details)

\begin{figure}
	\includegraphics[width=1\columnwidth]{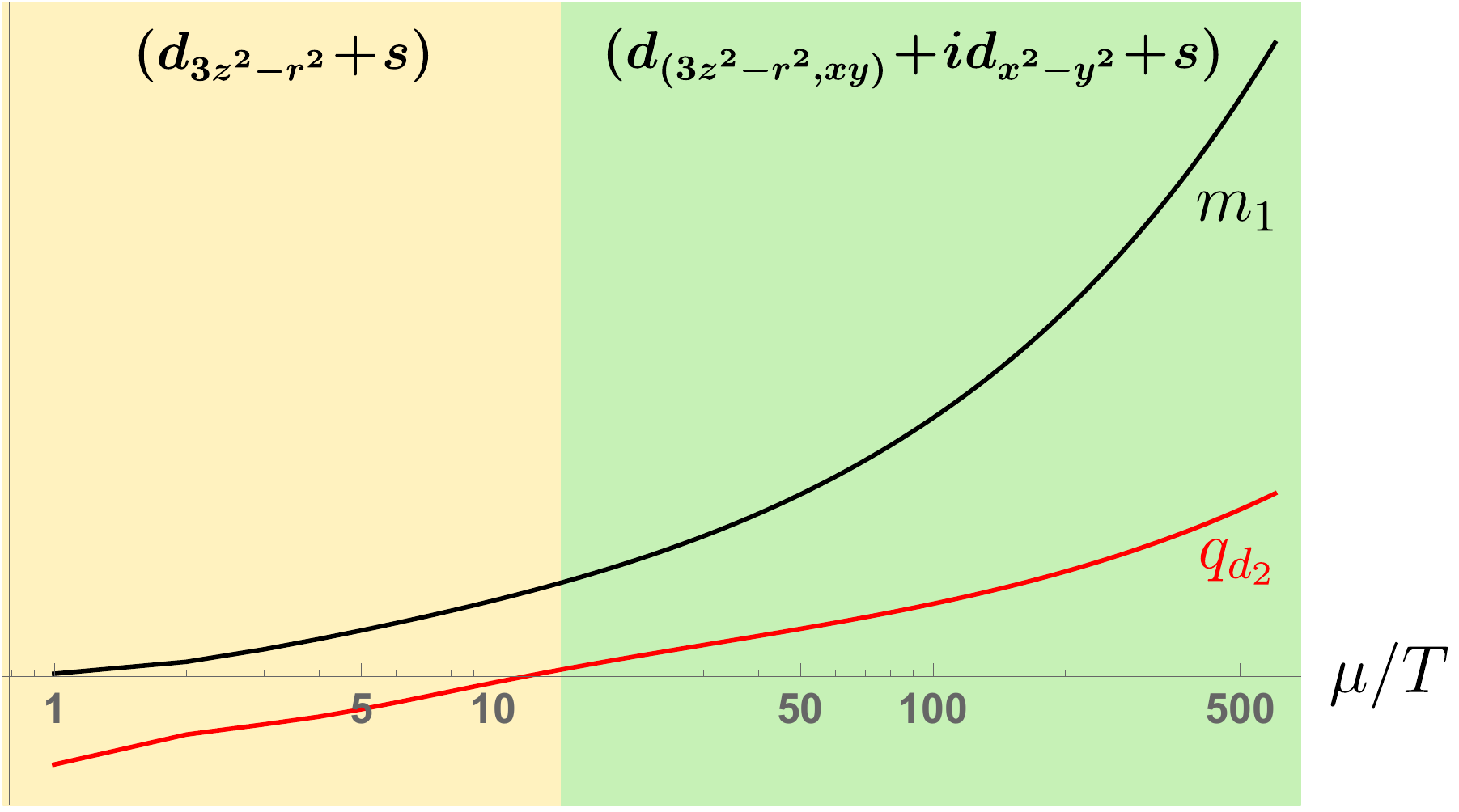}
	\caption{(color online) Plot of coefficients $m_1$ and $q_{d_2}$ as a function of $\mu/T$: Within one-loop calculation, using the parameter set for YPtBi, the values of $m_1$, $q_{d_2}$ and the corresponding superconducting phases are shown as a function of $\mu/T$. ($\mu$ is the chemical potential, and $T$ is temperature.) See main text for details.
	}
	\label{fig:lpdf}
\end{figure}

Finally, we briefly discuss application of our theory to relevant materials. In principle, our theoretical results can be generally used to investigate any systems, where the low energy kinetics is described by the Luttinger Hamiltonian. There are potential candidate materials which include lacunar spinel compounds GaM$_4$X$_8$, rare-earth cage compounds Pr(TM)${_2}$X${_{20}}$ and half-heusler compounds.\cite{abd2004transition,kim2014spin,matsumoto2016heavy,lin2010half,al2010topological} Among these candidates, we now focus on half-heusler compounds, which have  been widely discussed in recent years. Many half-heusler compounds such as YPtBi, TbPdBi, LuPdBi, LaPtBi exhibit superconductivity at low temperature.\cite{kim2018beyond,nakajima2015topological,goll2008thermodynamic} In these materials, metallic phase is well described by the Luttinger model having tiny Fermi pocket near ${\textbf{k}\!=\!0}$.\cite{butch2011superconductivity,meinert2016unconventional} Especially, YPtBi compounds become
superconducting below $T_c\!=\!0.77K$, which is relatively high transition temperature considering tiny electron density of this material, $n\!\approx\! 10^{19}\text{cm}^{-3}$.\cite{kim2018beyond,meinert2016unconventional}
Measurement on temperature dependence of penetration depth indicates that the superconducting order parameter may have s-wave contribution as well as a $T$-linear contribution\cite{kim2018beyond,roy2017topological}. This is quite an exotic phenomena, which may indicate the existence of gapless Bogoliubov quasiparticles. To understand the most promising superconducting phase, we adopt the parameter set that has been used to describe YPtBi; $\mu/\Lambda^2\!=\!0.4$ with a chemical potential $\mu$ and an ultraviolet cutoff $\Lambda$, $c_0\!=\!0.17$ and $c_a\!=\!1$ defined in Eq.\eqref{eq:h}.\cite{boettcher2018unconventional}
By performing one-loop calculations, we show how the values of $q_{d_2}$ and $m_1$ vary as a function of $\mu/T$ in Fig.\ref{fig:lpdf}.
Note that the coefficient $q_{d_3}$ for quartic invariant $\text{tr}((\phi^\dagger\phi)^2)$ vanishes within the one-loop expansion. (See SI for details.) Then, the phase transition between $(d_{3z^2-r^2}\!+\!s)$ and $(d_{(3z^2-r^2,xy)}\!+\!id_{x^2-y^2}\!+\!s)$ states occur at $\mu/T\!=\!15$. For YPtBi, quantum oscillation measurement reveals $\mu/T_c\!\simeq \!500$. Thus, within our theory, we speculate the system may favor topological superconductor with $(d_{(3z^2-r^2,xy)}\!+\!id_{x^2-y^2}\!+\!s)$, where gapless Bogoliubov quasiparticles form Fermi pockets having Chern numbers $\pm2$ as discussed above. Understanding the relation between symmetry analyses done in this work and any microscopic mechanism for superconductivity in half-heusler compounds remains as an outstanding issue. Further studies of the structures of topological defects in these topological superconductors
would also be interesting topics of future study.


\begin{acknowledgments}
We thank Igor F. Herbut, Hyunsoo Kim, Andrey V. Chubukov, Andriy H. Nevidomskyy, Lucile Savary,  Eun-Gook Moon, MyungJoon Han and Leon Balents for many useful discussions. 
Y.B.K is supported by the NSERC of Canada, Canadian Institute for Advanced
Research, and Center for Quantum Materials at the University of Toronto. G.Y.C is supported by BK21 plus program, POSTECH. A.M is supported by BK21 plus program, KAIST. 
S.B.L. is supported by the KAIST startup and National Research Foundation Grant (NRF-2017R1A2B4008097). 
\end{acknowledgments}

\bibliography{luttinger_super_bib}

\begin{thebibliography}{42}%
\makeatletter
\providecommand \@ifxundefined [1]{%
 \@ifx{#1\undefined}
}%
\providecommand \@ifnum [1]{%
 \ifnum #1\expandafter \@firstoftwo
 \else \expandafter \@secondoftwo
 \fi
}%
\providecommand \@ifx [1]{%
 \ifx #1\expandafter \@firstoftwo
 \else \expandafter \@secondoftwo
 \fi
}%
\providecommand \natexlab [1]{#1}%
\providecommand \enquote  [1]{``#1''}%
\providecommand \bibnamefont  [1]{#1}%
\providecommand \bibfnamefont [1]{#1}%
\providecommand \citenamefont [1]{#1}%
\providecommand \href@noop [0]{\@secondoftwo}%
\providecommand \href [0]{\begingroup \@sanitize@url \@href}%
\providecommand \@href[1]{\@@startlink{#1}\@@href}%
\providecommand \@@href[1]{\endgroup#1\@@endlink}%
\providecommand \@sanitize@url [0]{\catcode `\\12\catcode `\$12\catcode
  `\&12\catcode `\#12\catcode `\^12\catcode `\_12\catcode `\%12\relax}%
\providecommand \@@startlink[1]{}%
\providecommand \@@endlink[0]{}%
\providecommand \url  [0]{\begingroup\@sanitize@url \@url }%
\providecommand \@url [1]{\endgroup\@href {#1}{\urlprefix }}%
\providecommand \urlprefix  [0]{URL }%
\providecommand \Eprint [0]{\href }%
\providecommand \doibase [0]{http://dx.doi.org/}%
\providecommand \selectlanguage [0]{\@gobble}%
\providecommand \bibinfo  [0]{\@secondoftwo}%
\providecommand \bibfield  [0]{\@secondoftwo}%
\providecommand \translation [1]{[#1]}%
\providecommand \BibitemOpen [0]{}%
\providecommand \bibitemStop [0]{}%
\providecommand \bibitemNoStop [0]{.\EOS\space}%
\providecommand \EOS [0]{\spacefactor3000\relax}%
\providecommand \BibitemShut  [1]{\csname bibitem#1\endcsname}%
\let\auto@bib@innerbib\@empty
\bibitem [{\citenamefont {Fu}\ and\ \citenamefont
  {Kane}(2008)}]{fu2008superconducting}%
  \BibitemOpen
  \bibfield  {author} {\bibinfo {author} {\bibfnamefont {L.}~\bibnamefont
  {Fu}}\ and\ \bibinfo {author} {\bibfnamefont {C.~L.}\ \bibnamefont {Kane}},\
  }\href@noop {} {\bibfield  {journal} {\bibinfo  {journal} {Physical review
  letters}\ }\textbf {\bibinfo {volume} {100}},\ \bibinfo {pages} {096407}
  (\bibinfo {year} {2008})}\BibitemShut {NoStop}%
\bibitem [{\citenamefont {Hasan}\ and\ \citenamefont
  {Kane}(2010)}]{hasan2010colloquium}%
  \BibitemOpen
  \bibfield  {author} {\bibinfo {author} {\bibfnamefont {M.~Z.}\ \bibnamefont
  {Hasan}}\ and\ \bibinfo {author} {\bibfnamefont {C.~L.}\ \bibnamefont
  {Kane}},\ }\href@noop {} {\bibfield  {journal} {\bibinfo  {journal} {Reviews
  of Modern Physics}\ }\textbf {\bibinfo {volume} {82}},\ \bibinfo {pages}
  {3045} (\bibinfo {year} {2010})}\BibitemShut {NoStop}%
\bibitem [{\citenamefont {Qi}\ and\ \citenamefont
  {Zhang}(2011)}]{qi2011topological}%
  \BibitemOpen
  \bibfield  {author} {\bibinfo {author} {\bibfnamefont {X.-L.}\ \bibnamefont
  {Qi}}\ and\ \bibinfo {author} {\bibfnamefont {S.-C.}\ \bibnamefont {Zhang}},\
  }\href@noop {} {\bibfield  {journal} {\bibinfo  {journal} {Reviews of Modern
  Physics}\ }\textbf {\bibinfo {volume} {83}},\ \bibinfo {pages} {1057}
  (\bibinfo {year} {2011})}\BibitemShut {NoStop}%
\bibitem [{\citenamefont {Wan}\ \emph {et~al.}(2011)\citenamefont {Wan},
  \citenamefont {Turner}, \citenamefont {Vishwanath},\ and\ \citenamefont
  {Savrasov}}]{wan2011topological}%
  \BibitemOpen
  \bibfield  {author} {\bibinfo {author} {\bibfnamefont {X.}~\bibnamefont
  {Wan}}, \bibinfo {author} {\bibfnamefont {A.~M.}\ \bibnamefont {Turner}},
  \bibinfo {author} {\bibfnamefont {A.}~\bibnamefont {Vishwanath}}, \ and\
  \bibinfo {author} {\bibfnamefont {S.~Y.}\ \bibnamefont {Savrasov}},\
  }\href@noop {} {\bibfield  {journal} {\bibinfo  {journal} {Physical Review
  B}\ }\textbf {\bibinfo {volume} {83}},\ \bibinfo {pages} {205101} (\bibinfo
  {year} {2011})}\BibitemShut {NoStop}%
\bibitem [{\citenamefont {Burkov}\ \emph {et~al.}(2011)\citenamefont {Burkov},
  \citenamefont {Hook},\ and\ \citenamefont {Balents}}]{burkov2011topological}%
  \BibitemOpen
  \bibfield  {author} {\bibinfo {author} {\bibfnamefont {A.}~\bibnamefont
  {Burkov}}, \bibinfo {author} {\bibfnamefont {M.}~\bibnamefont {Hook}}, \ and\
  \bibinfo {author} {\bibfnamefont {L.}~\bibnamefont {Balents}},\ }\href@noop
  {} {\bibfield  {journal} {\bibinfo  {journal} {Physical Review B}\ }\textbf
  {\bibinfo {volume} {84}},\ \bibinfo {pages} {235126} (\bibinfo {year}
  {2011})}\BibitemShut {NoStop}%
\bibitem [{\citenamefont {Chiu}\ \emph {et~al.}(2016)\citenamefont {Chiu},
  \citenamefont {Teo}, \citenamefont {Schnyder},\ and\ \citenamefont
  {Ryu}}]{chiu2016classification}%
  \BibitemOpen
  \bibfield  {author} {\bibinfo {author} {\bibfnamefont {C.-K.}\ \bibnamefont
  {Chiu}}, \bibinfo {author} {\bibfnamefont {J.~C.}\ \bibnamefont {Teo}},
  \bibinfo {author} {\bibfnamefont {A.~P.}\ \bibnamefont {Schnyder}}, \ and\
  \bibinfo {author} {\bibfnamefont {S.}~\bibnamefont {Ryu}},\ }\href@noop {}
  {\bibfield  {journal} {\bibinfo  {journal} {Reviews of Modern Physics}\
  }\textbf {\bibinfo {volume} {88}},\ \bibinfo {pages} {035005} (\bibinfo
  {year} {2016})}\BibitemShut {NoStop}%
\bibitem [{\citenamefont {Nayak}\ \emph {et~al.}(2008)\citenamefont {Nayak},
  \citenamefont {Simon}, \citenamefont {Stern}, \citenamefont {Freedman},\ and\
  \citenamefont {Sarma}}]{nayak2008non}%
  \BibitemOpen
  \bibfield  {author} {\bibinfo {author} {\bibfnamefont {C.}~\bibnamefont
  {Nayak}}, \bibinfo {author} {\bibfnamefont {S.~H.}\ \bibnamefont {Simon}},
  \bibinfo {author} {\bibfnamefont {A.}~\bibnamefont {Stern}}, \bibinfo
  {author} {\bibfnamefont {M.}~\bibnamefont {Freedman}}, \ and\ \bibinfo
  {author} {\bibfnamefont {S.~D.}\ \bibnamefont {Sarma}},\ }\href@noop {}
  {\bibfield  {journal} {\bibinfo  {journal} {Reviews of Modern Physics}\
  }\textbf {\bibinfo {volume} {80}},\ \bibinfo {pages} {1083} (\bibinfo {year}
  {2008})}\BibitemShut {NoStop}%
\bibitem [{\citenamefont {Lee}\ and\ \citenamefont {Wen}(2008)}]{lee2008spin}%
  \BibitemOpen
  \bibfield  {author} {\bibinfo {author} {\bibfnamefont {P.~A.}\ \bibnamefont
  {Lee}}\ and\ \bibinfo {author} {\bibfnamefont {X.-G.}\ \bibnamefont {Wen}},\
  }\href@noop {} {\bibfield  {journal} {\bibinfo  {journal} {Physical review
  B}\ }\textbf {\bibinfo {volume} {78}},\ \bibinfo {pages} {144517} (\bibinfo
  {year} {2008})}\BibitemShut {NoStop}%
\bibitem [{\citenamefont {Mackenzie}\ and\ \citenamefont
  {Maeno}(2003)}]{mackenzie2003superconductivity}%
  \BibitemOpen
  \bibfield  {author} {\bibinfo {author} {\bibfnamefont {A.~P.}\ \bibnamefont
  {Mackenzie}}\ and\ \bibinfo {author} {\bibfnamefont {Y.}~\bibnamefont
  {Maeno}},\ }\href@noop {} {\bibfield  {journal} {\bibinfo  {journal} {Reviews
  of Modern Physics}\ }\textbf {\bibinfo {volume} {75}},\ \bibinfo {pages}
  {657} (\bibinfo {year} {2003})}\BibitemShut {NoStop}%
\bibitem [{\citenamefont {Kuroki}\ \emph {et~al.}(2001)\citenamefont {Kuroki},
  \citenamefont {Arita},\ and\ \citenamefont {Aoki}}]{kuroki2001spin}%
  \BibitemOpen
  \bibfield  {author} {\bibinfo {author} {\bibfnamefont {K.}~\bibnamefont
  {Kuroki}}, \bibinfo {author} {\bibfnamefont {R.}~\bibnamefont {Arita}}, \
  and\ \bibinfo {author} {\bibfnamefont {H.}~\bibnamefont {Aoki}},\ }\href@noop
  {} {\bibfield  {journal} {\bibinfo  {journal} {Physical Review B}\ }\textbf
  {\bibinfo {volume} {63}},\ \bibinfo {pages} {094509} (\bibinfo {year}
  {2001})}\BibitemShut {NoStop}%
\bibitem [{\citenamefont {Sigrist}\ and\ \citenamefont
  {Ueda}(1991)}]{sigrist1991phenomenological}%
  \BibitemOpen
  \bibfield  {author} {\bibinfo {author} {\bibfnamefont {M.}~\bibnamefont
  {Sigrist}}\ and\ \bibinfo {author} {\bibfnamefont {K.}~\bibnamefont {Ueda}},\
  }\href@noop {} {\bibfield  {journal} {\bibinfo  {journal} {Reviews of Modern
  physics}\ }\textbf {\bibinfo {volume} {63}},\ \bibinfo {pages} {239}
  (\bibinfo {year} {1991})}\BibitemShut {NoStop}%
\bibitem [{\citenamefont {Sato}\ and\ \citenamefont
  {Ando}(2017)}]{sato2017topological}%
  \BibitemOpen
  \bibfield  {author} {\bibinfo {author} {\bibfnamefont {M.}~\bibnamefont
  {Sato}}\ and\ \bibinfo {author} {\bibfnamefont {Y.}~\bibnamefont {Ando}},\
  }\href@noop {} {\bibfield  {journal} {\bibinfo  {journal} {Reports on
  Progress in Physics}\ }\textbf {\bibinfo {volume} {80}},\ \bibinfo {pages}
  {076501} (\bibinfo {year} {2017})}\BibitemShut {NoStop}%
\bibitem [{\citenamefont {Moreo}\ \emph {et~al.}(2009)\citenamefont {Moreo},
  \citenamefont {Daghofer}, \citenamefont {Nicholson},\ and\ \citenamefont
  {Dagotto}}]{moreo2009interband}%
  \BibitemOpen
  \bibfield  {author} {\bibinfo {author} {\bibfnamefont {A.}~\bibnamefont
  {Moreo}}, \bibinfo {author} {\bibfnamefont {M.}~\bibnamefont {Daghofer}},
  \bibinfo {author} {\bibfnamefont {A.}~\bibnamefont {Nicholson}}, \ and\
  \bibinfo {author} {\bibfnamefont {E.}~\bibnamefont {Dagotto}},\ }\href@noop
  {} {\bibfield  {journal} {\bibinfo  {journal} {Physical Review B}\ }\textbf
  {\bibinfo {volume} {80}},\ \bibinfo {pages} {104507} (\bibinfo {year}
  {2009})}\BibitemShut {NoStop}%
\bibitem [{\citenamefont {Boettcher}\ and\ \citenamefont
  {Herbut}(2018)}]{boettcher2018unconventional}%
  \BibitemOpen
  \bibfield  {author} {\bibinfo {author} {\bibfnamefont {I.}~\bibnamefont
  {Boettcher}}\ and\ \bibinfo {author} {\bibfnamefont {I.~F.}\ \bibnamefont
  {Herbut}},\ }\href@noop {} {\bibfield  {journal} {\bibinfo  {journal}
  {Physical review letters}\ }\textbf {\bibinfo {volume} {120}},\ \bibinfo
  {pages} {057002} (\bibinfo {year} {2018})}\BibitemShut {NoStop}%
\bibitem [{\citenamefont {Venderbos}\ \emph {et~al.}(2017)\citenamefont
  {Venderbos}, \citenamefont {Savary}, \citenamefont {Ruhman}, \citenamefont
  {Lee},\ and\ \citenamefont {Fu}}]{venderbos2017pairing}%
  \BibitemOpen
  \bibfield  {author} {\bibinfo {author} {\bibfnamefont {J.~W.}\ \bibnamefont
  {Venderbos}}, \bibinfo {author} {\bibfnamefont {L.}~\bibnamefont {Savary}},
  \bibinfo {author} {\bibfnamefont {J.}~\bibnamefont {Ruhman}}, \bibinfo
  {author} {\bibfnamefont {P.~A.}\ \bibnamefont {Lee}}, \ and\ \bibinfo
  {author} {\bibfnamefont {L.}~\bibnamefont {Fu}},\ }\href@noop {} {\bibfield
  {journal} {\bibinfo  {journal} {arXiv preprint arXiv:1709.04487}\ } (\bibinfo
  {year} {2017})}\BibitemShut {NoStop}%
\bibitem [{\citenamefont {Nomoto}\ \emph {et~al.}(2016)\citenamefont {Nomoto},
  \citenamefont {Hattori},\ and\ \citenamefont
  {Ikeda}}]{nomoto2016classification}%
  \BibitemOpen
  \bibfield  {author} {\bibinfo {author} {\bibfnamefont {T.}~\bibnamefont
  {Nomoto}}, \bibinfo {author} {\bibfnamefont {K.}~\bibnamefont {Hattori}}, \
  and\ \bibinfo {author} {\bibfnamefont {H.}~\bibnamefont {Ikeda}},\
  }\href@noop {} {\bibfield  {journal} {\bibinfo  {journal} {Physical Review
  B}\ }\textbf {\bibinfo {volume} {94}},\ \bibinfo {pages} {174513} (\bibinfo
  {year} {2016})}\BibitemShut {NoStop}%
\bibitem [{\citenamefont {Yanase}(2016)}]{yanase2016nonsymmorphic}%
  \BibitemOpen
  \bibfield  {author} {\bibinfo {author} {\bibfnamefont {Y.}~\bibnamefont
  {Yanase}},\ }\href@noop {} {\bibfield  {journal} {\bibinfo  {journal}
  {Physical Review B}\ }\textbf {\bibinfo {volume} {94}},\ \bibinfo {pages}
  {174502} (\bibinfo {year} {2016})}\BibitemShut {NoStop}%
\bibitem [{\citenamefont {Luttinger}\ and\ \citenamefont
  {Kohn}(1955)}]{luttinger1955motion}%
  \BibitemOpen
  \bibfield  {author} {\bibinfo {author} {\bibfnamefont {J.~M.}\ \bibnamefont
  {Luttinger}}\ and\ \bibinfo {author} {\bibfnamefont {W.}~\bibnamefont
  {Kohn}},\ }\href@noop {} {\bibfield  {journal} {\bibinfo  {journal} {Physical
  Review}\ }\textbf {\bibinfo {volume} {97}},\ \bibinfo {pages} {869} (\bibinfo
  {year} {1955})}\BibitemShut {NoStop}%
\bibitem [{\citenamefont {Savary}\ \emph {et~al.}(2014)\citenamefont {Savary},
  \citenamefont {Moon},\ and\ \citenamefont {Balents}}]{savary2014new}%
  \BibitemOpen
  \bibfield  {author} {\bibinfo {author} {\bibfnamefont {L.}~\bibnamefont
  {Savary}}, \bibinfo {author} {\bibfnamefont {E.-G.}\ \bibnamefont {Moon}}, \
  and\ \bibinfo {author} {\bibfnamefont {L.}~\bibnamefont {Balents}},\
  }\href@noop {} {\bibfield  {journal} {\bibinfo  {journal} {Physical Review
  X}\ }\textbf {\bibinfo {volume} {4}},\ \bibinfo {pages} {041027} (\bibinfo
  {year} {2014})}\BibitemShut {NoStop}%
\bibitem [{\citenamefont {Moon}\ \emph {et~al.}(2013)\citenamefont {Moon},
  \citenamefont {Xu}, \citenamefont {Kim},\ and\ \citenamefont
  {Balents}}]{moon2013non}%
  \BibitemOpen
  \bibfield  {author} {\bibinfo {author} {\bibfnamefont {E.-G.}\ \bibnamefont
  {Moon}}, \bibinfo {author} {\bibfnamefont {C.}~\bibnamefont {Xu}}, \bibinfo
  {author} {\bibfnamefont {Y.~B.}\ \bibnamefont {Kim}}, \ and\ \bibinfo
  {author} {\bibfnamefont {L.}~\bibnamefont {Balents}},\ }\href@noop {}
  {\bibfield  {journal} {\bibinfo  {journal} {Physical review letters}\
  }\textbf {\bibinfo {volume} {111}},\ \bibinfo {pages} {206401} (\bibinfo
  {year} {2013})}\BibitemShut {NoStop}%
\bibitem [{\citenamefont {Boettcher}\ and\ \citenamefont
  {Herbut}(2017)}]{boettcher2017anisotropy}%
  \BibitemOpen
  \bibfield  {author} {\bibinfo {author} {\bibfnamefont {I.}~\bibnamefont
  {Boettcher}}\ and\ \bibinfo {author} {\bibfnamefont {I.~F.}\ \bibnamefont
  {Herbut}},\ }\href@noop {} {\bibfield  {journal} {\bibinfo  {journal}
  {Physical Review B}\ }\textbf {\bibinfo {volume} {95}},\ \bibinfo {pages}
  {075149} (\bibinfo {year} {2017})}\BibitemShut {NoStop}%
\bibitem [{\citenamefont {Yang}\ and\ \citenamefont
  {Kim}(2010)}]{yang2010topological}%
  \BibitemOpen
  \bibfield  {author} {\bibinfo {author} {\bibfnamefont {B.-J.}\ \bibnamefont
  {Yang}}\ and\ \bibinfo {author} {\bibfnamefont {Y.~B.}\ \bibnamefont {Kim}},\
  }\href@noop {} {\bibfield  {journal} {\bibinfo  {journal} {Physical Review
  B}\ }\textbf {\bibinfo {volume} {82}},\ \bibinfo {pages} {085111} (\bibinfo
  {year} {2010})}\BibitemShut {NoStop}%
\bibitem [{\citenamefont {Tinkham}(2004)}]{tinkham2004introduction}%
  \BibitemOpen
  \bibfield  {author} {\bibinfo {author} {\bibfnamefont {M.}~\bibnamefont
  {Tinkham}},\ }\href@noop {} {\emph {\bibinfo {title} {Introduction to
  superconductivity}}}\ (\bibinfo  {publisher} {Courier Corporation},\ \bibinfo
  {year} {2004})\BibitemShut {NoStop}%
\bibitem [{\citenamefont {Agterberg}\ \emph {et~al.}(2017)\citenamefont
  {Agterberg}, \citenamefont {Brydon},\ and\ \citenamefont
  {Timm}}]{agterberg2017bogoliubov}%
  \BibitemOpen
  \bibfield  {author} {\bibinfo {author} {\bibfnamefont {D.}~\bibnamefont
  {Agterberg}}, \bibinfo {author} {\bibfnamefont {P.}~\bibnamefont {Brydon}}, \
  and\ \bibinfo {author} {\bibfnamefont {C.}~\bibnamefont {Timm}},\ }\href@noop
  {} {\bibfield  {journal} {\bibinfo  {journal} {Physical review letters}\
  }\textbf {\bibinfo {volume} {118}},\ \bibinfo {pages} {127001} (\bibinfo
  {year} {2017})}\BibitemShut {NoStop}%
\bibitem [{\citenamefont {Roy}\ \emph {et~al.}(2017)\citenamefont {Roy},
  \citenamefont {Ghorashi}, \citenamefont {Foster},\ and\ \citenamefont
  {Nevidomskyy}}]{roy2017topological}%
  \BibitemOpen
  \bibfield  {author} {\bibinfo {author} {\bibfnamefont {B.}~\bibnamefont
  {Roy}}, \bibinfo {author} {\bibfnamefont {S.~A.~A.}\ \bibnamefont
  {Ghorashi}}, \bibinfo {author} {\bibfnamefont {M.~S.}\ \bibnamefont
  {Foster}}, \ and\ \bibinfo {author} {\bibfnamefont {A.~H.}\ \bibnamefont
  {Nevidomskyy}},\ }\href@noop {} {\bibfield  {journal} {\bibinfo  {journal}
  {arXiv preprint arXiv:1708.07825}\ } (\bibinfo {year} {2017})}\BibitemShut
  {NoStop}%
\bibitem [{\citenamefont {Yu}\ and\ \citenamefont {Liu}(2018)}]{yu2018singlet}%
  \BibitemOpen
  \bibfield  {author} {\bibinfo {author} {\bibfnamefont {J.}~\bibnamefont
  {Yu}}\ and\ \bibinfo {author} {\bibfnamefont {C.-X.}\ \bibnamefont {Liu}},\
  }\href@noop {} {\bibfield  {journal} {\bibinfo  {journal} {Physical Review
  B}\ }\textbf {\bibinfo {volume} {98}},\ \bibinfo {pages} {104514} (\bibinfo
  {year} {2018})}\BibitemShut {NoStop}%
\bibitem [{\citenamefont {De~Matteis}\ \emph {et~al.}(2008)\citenamefont
  {De~Matteis}, \citenamefont {Sonnet},\ and\ \citenamefont
  {Virga}}]{de2008landau}%
  \BibitemOpen
  \bibfield  {author} {\bibinfo {author} {\bibfnamefont {G.}~\bibnamefont
  {De~Matteis}}, \bibinfo {author} {\bibfnamefont {A.~M.}\ \bibnamefont
  {Sonnet}}, \ and\ \bibinfo {author} {\bibfnamefont {E.~G.}\ \bibnamefont
  {Virga}},\ }\href@noop {} {\bibfield  {journal} {\bibinfo  {journal}
  {Continuum Mechanics and Thermodynamics}\ }\textbf {\bibinfo {volume} {20}},\
  \bibinfo {pages} {347} (\bibinfo {year} {2008})}\BibitemShut {NoStop}%
\bibitem [{\citenamefont {Brydon}\ \emph {et~al.}(2018)\citenamefont {Brydon},
  \citenamefont {Agterberg}, \citenamefont {Menke},\ and\ \citenamefont
  {Timm}}]{brydon2018bogoliubov}%
  \BibitemOpen
  \bibfield  {author} {\bibinfo {author} {\bibfnamefont {P.}~\bibnamefont
  {Brydon}}, \bibinfo {author} {\bibfnamefont {D.}~\bibnamefont {Agterberg}},
  \bibinfo {author} {\bibfnamefont {H.}~\bibnamefont {Menke}}, \ and\ \bibinfo
  {author} {\bibfnamefont {C.}~\bibnamefont {Timm}},\ }\href@noop {} {\bibfield
   {journal} {\bibinfo  {journal} {arXiv preprint arXiv:1806.03773}\ }
  (\bibinfo {year} {2018})}\BibitemShut {NoStop}%
\bibitem [{\citenamefont {Bzdu{\v{s}}ek}\ and\ \citenamefont
  {Sigrist}(2017)}]{bzduvsek2017robust}%
  \BibitemOpen
  \bibfield  {author} {\bibinfo {author} {\bibfnamefont {T.}~\bibnamefont
  {Bzdu{\v{s}}ek}}\ and\ \bibinfo {author} {\bibfnamefont {M.}~\bibnamefont
  {Sigrist}},\ }\href@noop {} {\bibfield  {journal} {\bibinfo  {journal}
  {Physical Review B}\ }\textbf {\bibinfo {volume} {96}},\ \bibinfo {pages}
  {155105} (\bibinfo {year} {2017})}\BibitemShut {NoStop}%
\bibitem [{\citenamefont {Brydon}\ \emph {et~al.}(2016)\citenamefont {Brydon},
  \citenamefont {Wang}, \citenamefont {Weinert},\ and\ \citenamefont
  {Agterberg}}]{brydon2016pairing}%
  \BibitemOpen
  \bibfield  {author} {\bibinfo {author} {\bibfnamefont {P.}~\bibnamefont
  {Brydon}}, \bibinfo {author} {\bibfnamefont {L.}~\bibnamefont {Wang}},
  \bibinfo {author} {\bibfnamefont {M.}~\bibnamefont {Weinert}}, \ and\
  \bibinfo {author} {\bibfnamefont {D.}~\bibnamefont {Agterberg}},\ }\href@noop
  {} {\bibfield  {journal} {\bibinfo  {journal} {Physical review letters}\
  }\textbf {\bibinfo {volume} {116}},\ \bibinfo {pages} {177001} (\bibinfo
  {year} {2016})}\BibitemShut {NoStop}%
\bibitem [{\citenamefont {Schnyder}\ and\ \citenamefont
  {Brydon}(2015)}]{schnyder2015topological}%
  \BibitemOpen
  \bibfield  {author} {\bibinfo {author} {\bibfnamefont {A.~P.}\ \bibnamefont
  {Schnyder}}\ and\ \bibinfo {author} {\bibfnamefont {P.~M.}\ \bibnamefont
  {Brydon}},\ }\href@noop {} {\bibfield  {journal} {\bibinfo  {journal}
  {Journal of Physics: Condensed Matter}\ }\textbf {\bibinfo {volume} {27}},\
  \bibinfo {pages} {243201} (\bibinfo {year} {2015})}\BibitemShut {NoStop}%
\bibitem [{\citenamefont {Abd-Elmeguid}\ \emph {et~al.}(2004)\citenamefont
  {Abd-Elmeguid}, \citenamefont {Ni}, \citenamefont {Khomskii}, \citenamefont
  {Pocha}, \citenamefont {Johrendt}, \citenamefont {Wang},\ and\ \citenamefont
  {Syassen}}]{abd2004transition}%
  \BibitemOpen
  \bibfield  {author} {\bibinfo {author} {\bibfnamefont {M.}~\bibnamefont
  {Abd-Elmeguid}}, \bibinfo {author} {\bibfnamefont {B.}~\bibnamefont {Ni}},
  \bibinfo {author} {\bibfnamefont {D.}~\bibnamefont {Khomskii}}, \bibinfo
  {author} {\bibfnamefont {R.}~\bibnamefont {Pocha}}, \bibinfo {author}
  {\bibfnamefont {D.}~\bibnamefont {Johrendt}}, \bibinfo {author}
  {\bibfnamefont {X.}~\bibnamefont {Wang}}, \ and\ \bibinfo {author}
  {\bibfnamefont {K.}~\bibnamefont {Syassen}},\ }\href@noop {} {\bibfield
  {journal} {\bibinfo  {journal} {Physical review letters}\ }\textbf {\bibinfo
  {volume} {93}},\ \bibinfo {pages} {126403} (\bibinfo {year}
  {2004})}\BibitemShut {NoStop}%
\bibitem [{\citenamefont {Kim}\ \emph {et~al.}(2014)\citenamefont {Kim},
  \citenamefont {Im}, \citenamefont {Han},\ and\ \citenamefont
  {Jin}}]{kim2014spin}%
  \BibitemOpen
  \bibfield  {author} {\bibinfo {author} {\bibfnamefont {H.-S.}\ \bibnamefont
  {Kim}}, \bibinfo {author} {\bibfnamefont {J.}~\bibnamefont {Im}}, \bibinfo
  {author} {\bibfnamefont {M.~J.}\ \bibnamefont {Han}}, \ and\ \bibinfo
  {author} {\bibfnamefont {H.}~\bibnamefont {Jin}},\ }\href@noop {} {\bibfield
  {journal} {\bibinfo  {journal} {Nature communications}\ }\textbf {\bibinfo
  {volume} {5}},\ \bibinfo {pages} {3988} (\bibinfo {year} {2014})}\BibitemShut
  {NoStop}%
\bibitem [{\citenamefont {Matsumoto}\ \emph {et~al.}(2016)\citenamefont
  {Matsumoto}, \citenamefont {Tsujimoto}, \citenamefont {Tomita}, \citenamefont
  {Sakai},\ and\ \citenamefont {Nakatsuji}}]{matsumoto2016heavy}%
  \BibitemOpen
  \bibfield  {author} {\bibinfo {author} {\bibfnamefont {Y.}~\bibnamefont
  {Matsumoto}}, \bibinfo {author} {\bibfnamefont {M.}~\bibnamefont
  {Tsujimoto}}, \bibinfo {author} {\bibfnamefont {T.}~\bibnamefont {Tomita}},
  \bibinfo {author} {\bibfnamefont {A.}~\bibnamefont {Sakai}}, \ and\ \bibinfo
  {author} {\bibfnamefont {S.}~\bibnamefont {Nakatsuji}},\ }in\ \href@noop {}
  {\emph {\bibinfo {booktitle} {Journal of Physics: Conference Series}}},\
  Vol.\ \bibinfo {volume} {683}\ (\bibinfo {organization} {IOP Publishing},\
  \bibinfo {year} {2016})\ p.\ \bibinfo {pages} {012013}\BibitemShut {NoStop}%
\bibitem [{\citenamefont {Lin}\ \emph {et~al.}(2010)\citenamefont {Lin},
  \citenamefont {Wray}, \citenamefont {Xia}, \citenamefont {Xu}, \citenamefont
  {Jia}, \citenamefont {Cava}, \citenamefont {Bansil},\ and\ \citenamefont
  {Hasan}}]{lin2010half}%
  \BibitemOpen
  \bibfield  {author} {\bibinfo {author} {\bibfnamefont {H.}~\bibnamefont
  {Lin}}, \bibinfo {author} {\bibfnamefont {L.~A.}\ \bibnamefont {Wray}},
  \bibinfo {author} {\bibfnamefont {Y.}~\bibnamefont {Xia}}, \bibinfo {author}
  {\bibfnamefont {S.}~\bibnamefont {Xu}}, \bibinfo {author} {\bibfnamefont
  {S.}~\bibnamefont {Jia}}, \bibinfo {author} {\bibfnamefont {R.~J.}\
  \bibnamefont {Cava}}, \bibinfo {author} {\bibfnamefont {A.}~\bibnamefont
  {Bansil}}, \ and\ \bibinfo {author} {\bibfnamefont {M.~Z.}\ \bibnamefont
  {Hasan}},\ }\href@noop {} {\bibfield  {journal} {\bibinfo  {journal} {Nature
  materials}\ }\textbf {\bibinfo {volume} {9}},\ \bibinfo {pages} {546}
  (\bibinfo {year} {2010})}\BibitemShut {NoStop}%
\bibitem [{\citenamefont {Al-Sawai}\ \emph {et~al.}(2010)\citenamefont
  {Al-Sawai}, \citenamefont {Lin}, \citenamefont {Markiewicz}, \citenamefont
  {Wray}, \citenamefont {Xia}, \citenamefont {Xu}, \citenamefont {Hasan},\ and\
  \citenamefont {Bansil}}]{al2010topological}%
  \BibitemOpen
  \bibfield  {author} {\bibinfo {author} {\bibfnamefont {W.}~\bibnamefont
  {Al-Sawai}}, \bibinfo {author} {\bibfnamefont {H.}~\bibnamefont {Lin}},
  \bibinfo {author} {\bibfnamefont {R.}~\bibnamefont {Markiewicz}}, \bibinfo
  {author} {\bibfnamefont {L.}~\bibnamefont {Wray}}, \bibinfo {author}
  {\bibfnamefont {Y.}~\bibnamefont {Xia}}, \bibinfo {author} {\bibfnamefont
  {S.-Y.}\ \bibnamefont {Xu}}, \bibinfo {author} {\bibfnamefont
  {M.}~\bibnamefont {Hasan}}, \ and\ \bibinfo {author} {\bibfnamefont
  {A.}~\bibnamefont {Bansil}},\ }\href@noop {} {\bibfield  {journal} {\bibinfo
  {journal} {Physical Review B}\ }\textbf {\bibinfo {volume} {82}},\ \bibinfo
  {pages} {125208} (\bibinfo {year} {2010})}\BibitemShut {NoStop}%
\bibitem [{\citenamefont {Kim}\ \emph {et~al.}(2018)\citenamefont {Kim},
  \citenamefont {Wang}, \citenamefont {Nakajima}, \citenamefont {Hu},
  \citenamefont {Ziemak}, \citenamefont {Syers}, \citenamefont {Wang},
  \citenamefont {Hodovanets}, \citenamefont {Denlinger}, \citenamefont {Brydon}
  \emph {et~al.}}]{kim2018beyond}%
  \BibitemOpen
  \bibfield  {author} {\bibinfo {author} {\bibfnamefont {H.}~\bibnamefont
  {Kim}}, \bibinfo {author} {\bibfnamefont {K.}~\bibnamefont {Wang}}, \bibinfo
  {author} {\bibfnamefont {Y.}~\bibnamefont {Nakajima}}, \bibinfo {author}
  {\bibfnamefont {R.}~\bibnamefont {Hu}}, \bibinfo {author} {\bibfnamefont
  {S.}~\bibnamefont {Ziemak}}, \bibinfo {author} {\bibfnamefont
  {P.}~\bibnamefont {Syers}}, \bibinfo {author} {\bibfnamefont
  {L.}~\bibnamefont {Wang}}, \bibinfo {author} {\bibfnamefont {H.}~\bibnamefont
  {Hodovanets}}, \bibinfo {author} {\bibfnamefont {J.~D.}\ \bibnamefont
  {Denlinger}}, \bibinfo {author} {\bibfnamefont {P.~M.}\ \bibnamefont
  {Brydon}},  \emph {et~al.},\ }\href@noop {} {\bibfield  {journal} {\bibinfo
  {journal} {Science advances}\ }\textbf {\bibinfo {volume} {4}},\ \bibinfo
  {pages} {eaao4513} (\bibinfo {year} {2018})}\BibitemShut {NoStop}%
\bibitem [{\citenamefont {Nakajima}\ \emph {et~al.}(2015)\citenamefont
  {Nakajima}, \citenamefont {Hu}, \citenamefont {Kirshenbaum}, \citenamefont
  {Hughes}, \citenamefont {Syers}, \citenamefont {Wang}, \citenamefont {Wang},
  \citenamefont {Wang}, \citenamefont {Saha}, \citenamefont {Pratt} \emph
  {et~al.}}]{nakajima2015topological}%
  \BibitemOpen
  \bibfield  {author} {\bibinfo {author} {\bibfnamefont {Y.}~\bibnamefont
  {Nakajima}}, \bibinfo {author} {\bibfnamefont {R.}~\bibnamefont {Hu}},
  \bibinfo {author} {\bibfnamefont {K.}~\bibnamefont {Kirshenbaum}}, \bibinfo
  {author} {\bibfnamefont {A.}~\bibnamefont {Hughes}}, \bibinfo {author}
  {\bibfnamefont {P.}~\bibnamefont {Syers}}, \bibinfo {author} {\bibfnamefont
  {X.}~\bibnamefont {Wang}}, \bibinfo {author} {\bibfnamefont {K.}~\bibnamefont
  {Wang}}, \bibinfo {author} {\bibfnamefont {R.}~\bibnamefont {Wang}}, \bibinfo
  {author} {\bibfnamefont {S.~R.}\ \bibnamefont {Saha}}, \bibinfo {author}
  {\bibfnamefont {D.}~\bibnamefont {Pratt}},  \emph {et~al.},\ }\href@noop {}
  {\bibfield  {journal} {\bibinfo  {journal} {Science advances}\ }\textbf
  {\bibinfo {volume} {1}},\ \bibinfo {pages} {e1500242} (\bibinfo {year}
  {2015})}\BibitemShut {NoStop}%
\bibitem [{\citenamefont {Goll}\ \emph {et~al.}(2008)\citenamefont {Goll},
  \citenamefont {Marz}, \citenamefont {Hamann}, \citenamefont {Tomanic},
  \citenamefont {Grube}, \citenamefont {Yoshino},\ and\ \citenamefont
  {Takabatake}}]{goll2008thermodynamic}%
  \BibitemOpen
  \bibfield  {author} {\bibinfo {author} {\bibfnamefont {G.}~\bibnamefont
  {Goll}}, \bibinfo {author} {\bibfnamefont {M.}~\bibnamefont {Marz}}, \bibinfo
  {author} {\bibfnamefont {A.}~\bibnamefont {Hamann}}, \bibinfo {author}
  {\bibfnamefont {T.}~\bibnamefont {Tomanic}}, \bibinfo {author} {\bibfnamefont
  {K.}~\bibnamefont {Grube}}, \bibinfo {author} {\bibfnamefont
  {T.}~\bibnamefont {Yoshino}}, \ and\ \bibinfo {author} {\bibfnamefont
  {T.}~\bibnamefont {Takabatake}},\ }\href@noop {} {\bibfield  {journal}
  {\bibinfo  {journal} {Physica B: Condensed Matter}\ }\textbf {\bibinfo
  {volume} {403}},\ \bibinfo {pages} {1065} (\bibinfo {year}
  {2008})}\BibitemShut {NoStop}%
\bibitem [{\citenamefont {Butch}\ \emph {et~al.}(2011)\citenamefont {Butch},
  \citenamefont {Syers}, \citenamefont {Kirshenbaum}, \citenamefont {Hope},\
  and\ \citenamefont {Paglione}}]{butch2011superconductivity}%
  \BibitemOpen
  \bibfield  {author} {\bibinfo {author} {\bibfnamefont {N.~P.}\ \bibnamefont
  {Butch}}, \bibinfo {author} {\bibfnamefont {P.}~\bibnamefont {Syers}},
  \bibinfo {author} {\bibfnamefont {K.}~\bibnamefont {Kirshenbaum}}, \bibinfo
  {author} {\bibfnamefont {A.~P.}\ \bibnamefont {Hope}}, \ and\ \bibinfo
  {author} {\bibfnamefont {J.}~\bibnamefont {Paglione}},\ }\href@noop {}
  {\bibfield  {journal} {\bibinfo  {journal} {Physical Review B}\ }\textbf
  {\bibinfo {volume} {84}},\ \bibinfo {pages} {220504} (\bibinfo {year}
  {2011})}\BibitemShut {NoStop}%
\bibitem [{\citenamefont {Meinert}(2016)}]{meinert2016unconventional}%
  \BibitemOpen
  \bibfield  {author} {\bibinfo {author} {\bibfnamefont {M.}~\bibnamefont
  {Meinert}},\ }\href@noop {} {\bibfield  {journal} {\bibinfo  {journal}
  {Physical review letters}\ }\textbf {\bibinfo {volume} {116}},\ \bibinfo
  {pages} {137001} (\bibinfo {year} {2016})}\BibitemShut {NoStop}%
\bibitem [{\citenamefont {Qi}\ \emph {et~al.}(2010)\citenamefont {Qi},
  \citenamefont {Hughes},\ and\ \citenamefont {Zhang}}]{qi2010topological}%
  \BibitemOpen
  \bibfield  {author} {\bibinfo {author} {\bibfnamefont {X.-L.}\ \bibnamefont
  {Qi}}, \bibinfo {author} {\bibfnamefont {T.~L.}\ \bibnamefont {Hughes}}, \
  and\ \bibinfo {author} {\bibfnamefont {S.-C.}\ \bibnamefont {Zhang}},\
  }\href@noop {} {\bibfield  {journal} {\bibinfo  {journal} {Physical Review
  B}\ }\textbf {\bibinfo {volume} {81}},\ \bibinfo {pages} {134508} (\bibinfo
  {year} {2010})}\BibitemShut {NoStop}%
\end{thebibliography}%

\newpage

\onecolumngrid

\appendix

\section{Luttinger Hamiltonian for a quadratic band touching system}
\label{sec:model}

In this section, we give the explicit expressions of Gell-Mann matrices $M_a$ and gamma matrices $\gamma_a$ and also discuss the relation between our model and general Luttinger Hamiltonian. 
$J_i$, which gives $i$-th component of angular momentum of $j=\frac{3}{2}$ states, are represented as follows,
\begin{gather}
	J_x=\left(
	\begin{array}{cccc}
		0 & \frac{\sqrt{3}}{2} & 0 & 0 \\
		\frac{\sqrt{3}}{2} & 0 & 1 & 0 \\
		0 & 1 & 0 & \frac{\sqrt{3}}{2} \\
		0 & 0 & \frac{\sqrt{3}}{2} & 0 \\
	\end{array}
	\right),J_y=\left(
	\begin{array}{cccc}
		0 & -\frac{i\sqrt{3}}{2} & 0 & 0 \\
		\frac{i \sqrt{3}}{2} & 0 & -i & 0 \\
		0 & i & 0 & -\frac{i\sqrt{3}}{2} \\
		0 & 0 & \frac{i \sqrt{3}}{2} & 0 \\
	\end{array}
	\right),J_z=\left(
	\begin{array}{cccc}
		\frac{3}{2} & 0 & 0 & 0 \\
		0 & \frac{1}{2} & 0 & 0 \\
		0 & 0 & -\frac{1}{2} & 0 \\
		0 & 0 & 0 & -\frac{3}{2} \\
	\end{array}
	\right).
\end{gather}
The $3\times3$ real Gell-Mann matrices $M_a$ and $4\times4$ $\gamma_a$ matrices are given as below,
\bea
M_1=
\left(
\begin{array}{ccc}
	1 & 0 & 0 \\
	0 & -1 & 0 \\
	0 & 0 & 0 \\
\end{array}
\right),
~M_2=
\frac{1}{\sqrt{3}}\left(
\begin{array}{ccc}
	-1 & 0 & 0 \\
	0 & -1 & 0 \\
	0 & 0 & 2 \\
\end{array}
\right),
~M_3=
\left(
\begin{array}{ccc}
	0 & 0 & 1 \\
	0 & 0 & 0 \\
	1 & 0 & 0 \\
\end{array}
\right),
~M_4=
\left(
\begin{array}{ccc}
	0 & 0 & 0 \\
	0 & 0 & 1 \\
	0 & 1 & 0 \\
\end{array}
\right),
~M_5=
\left(
\begin{array}{ccc}
	0 & 1 & 0 \\
	1 & 0 & 0 \\
	0 & 0 & 0 \\
\end{array}
\right),
\eea
\bea
\gamma_1=\frac{{J_x}^2-{J_y}^2}{\sqrt{3}}, 
~~\gamma_2={J_z}^2-\frac{5}{4}, 
~~\gamma_3=\frac{J_zJ_x+J_xJ_z}{\sqrt{3}}, 
~~\gamma_4=\frac{J_yJ_z+J_zJ_y}{\sqrt{3}}, 
~~\gamma_5=\frac{J_xJ_y+J_yJ_x}{\sqrt{3}},
\label{eq:g}
\eea
where $\gamma_a$ matrices satisfy the Clifford algebra $\{\gamma_a, \gamma_b\}= 2 \delta_{ab}$. 
Then, $d_a(\boldsymbol{k})$ can be written as follows,
\bea
d_1(\boldsymbol{k})=\frac{\sqrt{3}}{2}(k_x^2-k_y^2),~
d_2(\boldsymbol{k})=\frac{1}{2}(3k_z^2-k^2),~
d_3(\boldsymbol{k})=\sqrt{3}k_zk_x,~ 
d_4(\boldsymbol{k})=\sqrt{3}k_yk_z,~ 
d_5(\boldsymbol{k})=\sqrt{3}k_xk_y.
\eea

The general form of the Luttinger Hamiltonian matrix considering cubic symmetry without inversion is given as,
\bea
\tilde{h}_0(\boldsymbol{k})&=&\alpha	k^2 + \beta \sum_{i}k_{i}^2J_{i}^2 + \gamma \sum_{i \neq j}k_ik_jJ_iJ_j + \delta \sum_{i}k_i(J_{i+1}J_{i}J_{i+1}-J_{i-1}J_{i}J_{i-1})-\mu, \nonumber
\\
\nonumber
&=&(\alpha+\frac{5}{4}\beta) k^2 + \frac{1}{2}(\beta+\gamma) \sum_{a=1}^5 d_a(\boldsymbol{k})\gamma_a + \frac{1}{2}(-\beta+\gamma) \sum_{a=1}^5 s_a d_a(\boldsymbol{k})\gamma_a + \delta \sum_{i}k_i(J_{i+1}J_{i}J_{i+1}-J_{i-1}J_{i}J_{i-1})-\mu,
\\
\label{eq:h}
\eea
with $s_{1,2}=-1$ and $s_{3,4,5}=+1$. We set $\hbar=k_B=2m^*=1$ where $m^*$ is the effective electron mass. Here $(\alpha+\frac{5}{4}\beta)$ and $\mu$ quantify particle-hole asymmetry, $(-\beta+\gamma)$ measures cubic-anisotropy, $\delta$ is related to inversion odd terms which should vanish for centrosymmetric system. When $\beta=\gamma$ and $\delta=0$, $\tilde{h}_0(\boldsymbol{k})$ describes $SO(3)$ symmetric system and is consistent with $h_{0}(\boldsymbol{k})$ introduced in the main text. Considering fully spherical symmetric system, we set $(\alpha+\frac{5}{4}\beta)=c_0$ and $(\beta+\gamma)=1$ from here. 

For a half-heusler material YPtBi, it has been estimated $\alpha=20.5\text{eV}a^2/\pi^2$, $\beta=-18.5\text{eV}a^2/\pi^2$, $\gamma=-12.7\text{eV}a^2/\pi^2$, $\delta=0.06\text{eV}a/\pi$ and $a$ is the lattice constant.\cite{kim2018beyond} Thus, one can consider the system is close to $SO(3)$ symmetric with small cubic anisotropy and negligible inversion symmetry breaking since $\beta \approx \gamma$ and $\delta\approx 0$. Furthermore, quantum oscillation and angle resolved photoemission spectroscopy measurement reveals $\mu=35\text{meV}$ and $\mu=300\text{meV}$, respectively. With $T_c$=0.77K and $\mu=35\text{meV}$, we estimate $\mu/T_c\approx500$. Following Ref.\onlinecite{boettcher2018unconventional}, we also estimate $\mu/\Lambda^2\approx0.4$ regarding the bandwidth, where $\Lambda$ is the ultraviolet cutoff.

The bare on site Coulomb interaction, on the other hand, is written as,
\bea
h_{\text{int}}=g_0(\psi^\dagger\psi)^2 + \sum_{a}g_a(\psi^\dagger\gamma_a\psi)^2.
\label{eq:h_int}
\eea
Here, Eq.\ref{eq:g} clearly shows that $\psi^\dagger\gamma_a\psi$ transforms as d-wave orbitals or quadrupolar moments. Using the Fierz identiy, Eq.\ref{eq:h_int} is exactly decomposed as $h_s+\sum_a h_{d_a}$ with
\bea
h_{\text{int},s}&=&g_s(\psi^\dagger\gamma_{45}\psi^*)(\psi^T\gamma_{45}\psi),\\ 
h_{\text{int},d_a}&=&g_{d_a}(\psi^\dagger\gamma_a\gamma_{45}\psi^*)(\psi^T\gamma_{45}\gamma_a\psi),
\eea
where
\bea
g_s&=&\frac{1}{4}(g_0+\sum_a g_a), \\
g_{d_a}&=&\frac{1}{4} (g_0+g_a-\sum_{b \neq a}g_b).
\eea 
For fully spherical symmetric case, $g_a=g_{1}$ which leads to
\bea
g_s&=&\frac{1}{4}(g_0+5g_1), \\
g_d&=&\frac{1}{4}(g_0-3g_1).
\eea
Here, $\gamma_{45}\equiv i \gamma_{4}\gamma_{5}$ and the time-reversal operator is represented as $\mathcal{T}=\gamma_{45}\mathcal{K}$, where $\mathcal{K}$ indicates complex conjugation.
Now we define order parameters, $\Delta_s\equiv\langle\psi^T\gamma_{45}\psi\rangle$ and $\Delta_a\equiv\langle\psi^T\gamma_{45}\gamma_a\psi\rangle$ which corresponds to s- and d-wave superconducting order parameters, respectively. 

\section{Ginzburg-Landau Free Energy and one-loop expansion}
\label{sec:glf}

\begin{figure}[H]
	\centering\includegraphics[width=0.4\textwidth]{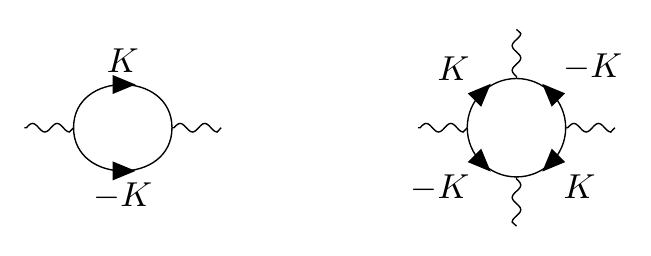}
	\caption{Diagrammatic representation of $L_{ab}$ and $L_{abcd}$. Each solid arrow refer free electron propagator, $G(K)$, with $K=(k_0,\boldsymbol{k})$ while each wiggly line indicate insertion of $\Delta_a$ with vertex $\gamma_a$.}
	\label{fig:fd}
\end{figure}

In this section, we show the Ginzburg-Landau free energy $F$ introduced in the main text. We consider the decomposed interactions 
and calculate the coefficients of $F$ within one-loop expansion upto quartic order. Before proceeding, we first introduce the free electron propagator as below,
\bea
G(K)=(ik_0 + c_0k^2+\sum_{a}d_a(\boldsymbol{k})\gamma_a-\mu)^{-1}=\frac{-ik_0-c_0k^2 + \sum_{a}d_a(\boldsymbol{k})\gamma_a+\mu}{(k_0-i((c_0+1)k^2-\mu))(k_0+i((-c_0+1)k^2+\mu))},
\eea
where $K\equiv(k_0,\boldsymbol{k})$ and $k_0=2\pi/T(n+1/2)$ indicates fermionic Matsubara frequency.
The free energy can be written as,
\bea
F(\Delta_s,\vec{\Delta})=\frac{1}{|g_s|}\Delta^*_s\Delta_s + \sum_a\frac{1}{|g_d|} \Delta^*_a\Delta_a + T\sum_{m,k_0}\int_{\boldsymbol{k}}^{\Lambda} \frac{1}{m} \text{tr}(-G(K)\Delta G(-K)^T\Delta^\dagger)^m,
\eea
where $\Delta=\gamma_{45}\Delta_s+\sum_a \gamma_a\gamma_{45}\Delta_a$. 
Let $F_{n}(\Delta_s,\vec{\Delta})$ be the gathering of an expansion of $F(\Delta_s,\vec{\Delta})$ which contains $n$-th power of $\Delta_s$ or $\Delta_a$. 
\bea
F_2(\Delta_s,\vec{\Delta})&=&\frac{1}{|g_s|}\Delta^*_s\Delta_s+\sum_a\frac{1}{|g_d|} \Delta^*_a\Delta_a -\frac{1}{2}\sum_{a,b}L_{ab}\Delta^*_{a}\Delta_{b},
\label{eq:f2}
\\
F_4(\Delta_s,\vec{\Delta})&=&\frac{1}{4}\sum_{a,b,c,d}L_{abcd}\Delta^*_{a}\Delta_{b}\Delta^*_{c}\Delta_{d},
\label{eq:f4}
\eea
with
\bea
L_{ab}&=&T\sum_{k_0}\int_{\boldsymbol{k}}^{\Lambda} \text{tr}(G(K)\gamma_aG(-K)\gamma_b),\\
L_{abcd}&=&T\sum_{k_0}\int_{\boldsymbol{k}}^{\Lambda} \text{tr}(G(K)\gamma_aG(-K)\gamma_bG(K)\gamma_cG(-K)\gamma_d),\\
\eea
which can be represented as diagrams shown in Fig.\ref{fig:fd}.

Meanwhile, the general $F_n(\Delta_s,\vec{\Delta})$ for fully symmetric system can be parametrized using the invariant theory,\cite{boettcher2018unconventional}
\bea
\nonumber
F_2(\Delta_s,\vec{\Delta})&=&r_d|\vec{\Delta}|^2+r_s|\Delta_s|^2
\\
\nonumber
F_4(\Delta_s,\vec{\Delta})&=&q_{d_1}|\vec{\Delta}|^4 +q_{d_2}|\vec{\Delta}^2|^2+q_{d_3}\text{tr}((\phi^\dagger\phi)^2) +q_{s}|\Delta_s|^4
+m_1(\text{tr}(\phi^2\phi^\dagger)\Delta_s^*+c.c.)+m_2(|\vec{\Delta}|^2|\Delta_s|^2) +m_3(\vec{\Delta}^2(\Delta_s^*)^2+c.c.)\\
\label{eq:free4}
\eea
with $\phi \equiv \sum_{a}\Delta_a M^a$. We take specific configurations of $\underbar{$\Delta$}^i\equiv(\Delta_s,\vec{\Delta})$ as below,
\bea
\nonumber 
\underbar{$\Delta$}^1&=&(0,0,1,0,0,0),
\underbar{$\Delta$}^2=\frac{1}{\sqrt{2}}(0,1,i,0,0,0),
\underbar{$\Delta$}^3=\frac{1}{\sqrt{2}}(0,0,0,1,i,0), \\
\underbar{$\Delta$}^4&=&\frac{1}{\sqrt{2}}(1,1,0,0,0,0),
\underbar{$\Delta$}^5=\frac{1}{\sqrt{2}}(1,0,1,0,0,0),
\underbar{$\Delta$}^6=\frac{1}{\sqrt{2}}(i,1,0,0,0,0),
\underbar{$\Delta$}^7=(1,0,0,0,0,0).
\label{eq:ansatz}
\eea
By applying \ref{eq:ansatz} into Eq.\ref{eq:f2}-\ref{eq:f4}, we get
\bea
F_2(\underbar{$\Delta$}^1)=r_d,
F_2(\underbar{$\Delta$}^7)=r_s
\eea
and
\bea
F_4(\underbar{$\Delta$}^1)&=&q_{d_1}+q_{d_2}+2q_{d_3},\\
F_4(\underbar{$\Delta$}^2)&=&q_{d_1}+\frac{4}{3}q_{d_3},\\
F_4(\underbar{$\Delta$}^3)&=&q_{d_1}+2q_{d_3},\\
F_4(\underbar{$\Delta$}^4)&=&\frac{1}{4}q_{d_1}+\frac{1}{4}q_{d_2}+\frac{1}{2}q_{d_3}+\frac{1}{2}q_s+\frac{1}{4}m_2+\frac{1}{2}m_3,\\
F_4(\underbar{$\Delta$}^5)&=&\frac{1}{4}q_{d_1}+\frac{1}{4}q_{d_2}+\frac{1}{2}q_{d_3}+\frac{1}{2}q_s+\frac{1}{\sqrt{3}}m_1+\frac{1}{4}m_2+\frac{1}{2}m_3,\\
F_4(\underbar{$\Delta$}^6)&=&\frac{1}{4}q_{d_1}+\frac{1}{4}q_{d_2}+\frac{1}{2}q_{d_3}+\frac{1}{2}q_s+\frac{1}{4}m_2-\frac{1}{2}m_3,\\
F_4(\underbar{$\Delta$}^7)&=&q_s.
\eea
Then one can extract relevant coefficients of the quadratic terms and quartic terms as below.
\bea
r_{d}&=&\frac{1}{|g_d|}|\vec{\Delta}|^2-T\sum_{k_0}\int_{\boldsymbol{k}}^{\Lambda}\frac{\frac{2}{5} \left(\left(5 c_0^2-3\right) k^4-10 c_0 k^2 \mu +5 \left(k_0^2+\mu ^2\right)\right)}{\left(\left(-c_0 k^2+k^2+\mu \right){}^2+k_0^2\right) \left(\left(c_0 k^2+k^2-\mu \right){}^2+k_0^2\right)},\\
r_{s}&=&\frac{1}{|g_s|}|\Delta_s|^2-T\sum_{k_0}\int_{\boldsymbol{k}}^{\Lambda}\frac{2 \left(\left(c_0^2+1\right) k^4-2 c_0 k^2 \mu +k_0^2+\mu ^2\right)}{\left(\left(-c_0 k^2+k^2+\mu \right){}^2+k_0^2\right) \left(\left(c_0 k^2+k^2-\mu \right){}^2+k_0^2\right)},\\
q_{d_1}&=&T\sum_{k_0}\int_{\boldsymbol{k}}^{\Lambda}\frac{g_1(k_0,\boldsymbol{k},\mu)}{\left(\left(-c_0 k^2+k^2+\mu \right){}^2+k_0^2\right){}^2 \left(\left(c_0 k^2+k^2-\mu \right){}^2+k_0^2\right){}^2},\\
q_{d_2}&=&T\sum_{k_0}\int_{\boldsymbol{k}}^{\Lambda}\frac{g_2(k_0,\boldsymbol{k},\mu)}{\left(\left(-c_0 k^2+k^2+\mu \right){}^2+k_0^2\right){}^2 \left(\left(c_0 k^2+k^2-\mu \right){}^2+k_0^2\right){}^2},\\
q_{s}&=&T\sum_{k_0}\int_{\boldsymbol{k}}^{\Lambda}\frac{g_3(k_0,\boldsymbol{k},\mu)}{\left(\left(-c_0 k^2+k^2+\mu \right){}^2+k_0^2\right){}^2 \left(\left(c_0 k^2+k^2-\mu \right){}^2+k_0^2\right){}^2},\\
m_{1}&=&T\sum_{k_0}\int_{\boldsymbol{k}}^{\Lambda}\frac{g_4(k_0,\boldsymbol{k},\mu)}{\left(\left(-c_0 k^2+k^2+\mu \right){}^2+k_0^2\right){}^2 \left(\left(c_0 k^2+k^2-\mu \right){}^2+k_0^2\right){}^2},\\
m_{2}&=&T\sum_{k_0}\int_{\boldsymbol{k}}^{\Lambda}\frac{g_5(k_0,\boldsymbol{k},\mu)}{\left(\left(-c_0 k^2+k^2+\mu \right){}^2+k_0^2\right){}^2 \left(\left(c_0 k^2+k^2-\mu \right){}^2+k_0^2\right){}^2},\\
m_{3}&=&T\sum_{k_0}\int_{\boldsymbol{k}}^{\Lambda}\frac{g_6(k_0,\boldsymbol{k},\mu)}{\left(\left(-c_0 k^2+k^2+\mu \right){}^2+k_0^2\right){}^2 \left(\left(c_0 k^2+k^2-\mu \right){}^2+k_0^2\right){}^2},\\
\eea
with
\bea
\nonumber
g_1(k_0,\boldsymbol{k},\mu)&=&\frac{2}{35}(-28k^6 \mu c_0(-3+5c_0^2)+k^8(15-42c_0^2+35c_0^4)-140k^2 \mu c_0(\mu^2+k_0^2)\\
&+&35(\mu^2+k_0^2)^2+14k^4(3\mu^2(-1+5c_0^2)+(-9+5c_0^2)k_0^2)),\\
\nonumber
g_2(k_0,\boldsymbol{k},\mu)&=&\frac{1}{35}(140k^6 \mu c_0(-1+c_0^2)-k^8(27-70c_0^2+35c_0^4)+140k^2 \mu c_0(\mu^2+k_0^2)\\
&-&35(\mu^2+k_0^2)^2-14k^4(5\mu^2(-1+3c_0^2)+(-7+5c_0^2)k_0^2)),\\
\nonumber
g_3(k_0,\boldsymbol{k},\mu)&=&-4k^6 \mu c_0(3+c_0^2)+k^8(1+6c_0^2+c_0^4)-4k^2 \mu c_0(\mu^2+k_0^2)\\
&+&(\mu^2+k_0^2)^2+2k^4(1+c_0^2)(3\mu^2+k_0^2),\\
g_4(k_0,\boldsymbol{k},\mu)&=&-\frac{16}{35} \sqrt{3} k^6 \left(c_0 k^2-\mu \right),\label{eq:ph}\\
\nonumber
g_5(k_0,\boldsymbol{k},\mu)&=&\frac{4}{5}(-4k^6 \mu c_0(3+5c_0^2)+k^8(-3+6c_0^2+5c_0^4)-20k^2 \mu c_0(\mu^2+k_0^2)\\
&+&5(\mu^2+k_0^2)^2+2k^4(1+5c_0^2)(3\mu^2+k_0^2)),\\
\nonumber
g_6(k_0,\boldsymbol{k},\mu)&=&\frac{1}{5}(4k^6 \mu c_0(1-5c_0^2)+k^8(5-2c_0^2+5c_0^4)-20k^2 \mu c_0(\mu^2+k_0^2)\\
&+&5(\mu^2+k_0^2)^2+2k^4(\mu^2(-1+15c_0^2)+5(1+c_0^2)k_0^2)).
\eea
In Eq.\ref{eq:ph}, one can clearly see that $m_1$ should vanish for fine-tuned particle-hole symmetric case i.e., $c_0=0$ and $\mu=0$, as stated in the main text. We also note $q_{d_3}=0$ within one-loop expansion regardless of the presence of particle-hole symmetry so we drop it from here.

\section{Free Energy analysis within real manifolds of superconducting order parameters} 
\label{sec:fea}
Within real manifolds of superconducting order parameters, one can find $R$ such that $R\phi'R^T=\phi$ where $R \in {SO(3)}$, $\phi'=\sum_{a=1-5}\Delta'_aM^a$ and $\phi=\sum_{a=1,2}\Delta_aM^a$ for general $\phi'$. Here $\phi'$ and $\phi$ represent physically equivalent order parameters of the system, which respect global ${SO(3)}$ symmetry. So we can write down our free energy as below without loss of generality within real manifolds.
\bea
\nonumber
F(\Delta_s,\phi)&=&\frac{r_d}{2}\tr(\phi^2) +r_s\Delta_s^2 + \frac{q_{d_1}}{4}(\tr(\phi^2))^2 + \frac{q_{d_2}}{4}(\tr(\phi^2))^2 +q_{s}\Delta_s^4\\
&+&2m_1(\tr(\phi^3)\Delta_s)+\frac{m_2}{2}(\tr(\phi^2)\Delta_s^2) +m_3(\tr(\phi^2)\Delta_s^2)\\
&=&\tr(\phi^2)(\frac{r_d}{2}+\frac{m_2}{2}\Delta_s^2+m_3\Delta_s^2)+\tr(\phi^3)(2m_1\Delta_s)+(\tr(\phi^2))^2(\frac{q_{d_1}+q_{d_2}}{4})+r_s\Delta_s^2+q_s\Delta_s^4\\
&\equiv& \tr(\phi^2)a_1+\tr(\phi^3)a_2+(\tr(\phi^2))^2a_3+r_s\Delta_s^2+q_s\Delta_s^4.
\label{eq:freer}
\eea
Equilibrium condition of Eq.\eqref{eq:freer} is
\bea
D[F,\phi]=0 \iff a_2\tr(\phi^2)1_3=3a_2\phi^2+2a_1\phi+4a_3\tr(\phi^2)\phi
\label{eq:eqc1}
\eea
where $D$ denotes tensorial differentiation and $1_3$ is $3\times3$ identity matrix.\cite{de2008landau}
Physically distinct solutions of Eq.\eqref{eq:eqc1} are given by uniaxial nematic state,
\bea
\Delta_1=0, \Delta_2=\frac{-\sqrt{3}(2m_1\Delta_s)\pm\sqrt{3(2m_1\Delta_s)^2-16(\frac{r_d}{2}+\frac{m_2}{2}\Delta_s^2+m_3\Delta_s^2)(q_{d_1}+q_{d_2})}}{4(q_{d_1}+q_{d_2})}.
\label{eq:eqc1s}
\eea
Another equilibrium condition is given by,
\bea
\frac{\partial F}{\partial \Delta_s}=0 \iff 4q_{s}\Delta_s^3+((m_2+2m_3)\tr(\phi^2)+2r_s)\Delta_s+2m_1\tr(\phi^3)=0,
\label{eq:eqc2}
\eea 
which clearly shows that $\Delta_1=0$ and non-zero $m_1$ and $\Delta_2$ lead to the emergence of parasitic s-wave pairing $\Delta_s\neq0$.
Substituting $\Delta_2$ using Eq.\ref{eq:eqc1s} gives us the exact value of $\Delta_s$, $\Delta_2$ and $F$ for a given parameter set, $(r_d,r_s,q_{d_1},q_{d_2},q_s,m_1,m_2,m_3)$.

Within complex manifolds, there exist $R$ and $\alpha$ such that $\Delta_s'e^{i\alpha}=\Delta_s$ and $R\phi' R^Te^{i\alpha}=\phi$ where $R \in {SO(3)}$, $\alpha \in \mathbb{R}$, $\phi'=\sum_{a=1-5}\Delta'_aM^a$ and $\phi=\sum_{a=1-5}\Delta_aM^a$ for general $\phi'$. Here, $(\Delta_s',\Delta_1',\Delta_2',\Delta_3',\Delta_4',\Delta_5') \in \mathbb{C}^6$, $(\Delta_1,\Delta_2) \in \mathbb{C}^2$ and $(\Delta_s,\Delta_3,\Delta_4,\Delta_5) \in \mathbb{R}^4$ as stated in the main text.  

\section{Gap structure and Pfaffian calculation} 
\label{sec:gs}
In order to investigate the gap structure, we first look into our Bogoliubov-de Gennes (BdG) Hamiltonian,
\bea
\mathcal{H} = \sum_{\boldsymbol{k}}\Psi_{\boldsymbol{k}}^{\dagger}\mathcal{H}(\boldsymbol{k})\Psi_{\boldsymbol{k}},
\label{H}
\eea
in the Nambu spinor basis with $\Psi_{\boldsymbol{k}}=(\psi_{\boldsymbol{k}}^T,\psi_{\boldsymbol{-k}}^{\dagger})^T$. 
The coefficient matrix is given by,
\bea
\mathcal{H}(\boldsymbol{k})=\left(
\begin{array}{cc}
	h_0(\boldsymbol{k}) & \Delta \\
	\Delta^{\dagger} & -h_0^{T}(-\boldsymbol{k}) \\
\end{array}
\right),
\label{eq:Hb}
\eea
where $\Delta=\gamma_{45}\Delta_s+\sum_a \gamma_a\gamma_{45}\Delta_a$. The BdG Hamiltonian possesses both inversion symmetry $\mathcal{P}$ and particle-hole symmetry $\mathcal{C}$. $\mathcal{P}$ is defined as $\mathcal{P}\mathcal{H}(\boldsymbol{-k})\mathcal{P}^\dagger=\mathcal{H}(\boldsymbol{k})$ with $\mathcal{P}=\tau_0 1_4$, where $\tau_0$ is $2\times 2$ identity matrix which acts on particle-hole space and $1_4$ is $4\times4$ identity matrix. $\mathcal{C}$ is defined as $-\mathcal{C}\mathcal{H}^*(\boldsymbol{-k})\mathcal{C}^\dagger=\mathcal{H}(\boldsymbol{k})$ with $\mathcal{C}=\tau_x 1_4$ where $\tau_i$ are Pauli matrices. The product of these symmetries gives $-(\mathcal{C}\mathcal{P}^*)\mathcal{H}^*(\boldsymbol{k})(\mathcal{C}\mathcal{P}^*)^\dagger=\mathcal{H}(\boldsymbol{k})$ which implies $(\mathcal{C}\mathcal{P})^2=1$. Then we can define Pfaffian $P(\boldsymbol{k})$ that satisfies the relation $P^2(\boldsymbol{k})=\text{det}\mathcal{H}(\boldsymbol{k})$. This relation indicates that the zeros of $P({\boldsymbol{k}})$ give nodal points of $\mathcal{H}(\boldsymbol{k})$. $P(\boldsymbol{k})$ is represented as,
\bea
\nonumber P(\boldsymbol{k})=(\langle \underbar{$d$}, \underbar{$d$}  \rangle - \langle \underbar{$\Delta$}, \underbar{$\Delta$}^* \rangle )^2 +4|\langle \underbar{$d$}, \underbar{$\Delta$}  \rangle|^2 \\ + \langle \underbar{$\Delta$}, \underbar{$\Delta$}  \rangle \langle \underbar{$\Delta$}^*, \underbar{$\Delta$}^*  \rangle -\langle \underbar{$\Delta$}, \underbar{$\Delta$}^*  \rangle ^2,
\label{eq:Pf}
\eea
where $\underbar{$d$} \equiv (c_0k^2-\mu,\vec{d}(\boldsymbol{k})), \underbar{$\Delta$} \equiv (\Delta_s,\vec{\Delta})$ and $\langle \underbar a ,\underbar b \rangle \equiv a_0b_0-\vec{a}\cdot\vec{b}$.\cite{brydon2016pairing} 
For the superconducting phase with ($d_{3z^2-r^2}\!+\!s$) pairing, $P(\boldsymbol{k})$ is given as,
\bea
P(\boldsymbol{k})=((d_0^2-k^4)-(\Delta_s^2-\Delta_2^2))^2+4(d_0\Delta_s-d_2\Delta_2)^2,
\label{eq:Pfun}
\eea
where $d_0=(c_0k^2-\mu)$.
So it cannot be a negative value for any $k$-points. But it vanishes where below two equalities are both satisfied.
\bea
(d_0^2-k^4)-(\Delta_s^2-\Delta_2^2)=0,
\label{eq:rpf1}
\\
d_0\Delta_s-d_2\Delta_2=0,
\label{eq:rpf2}
\eea
Indeed, the energy spectrum of the BdG Hamiltonian with ($d_{3z^2-r^2}\!+\!s$) pairing is given by,
\bea
|E_{\pm}(\boldsymbol{k})|=\sqrt{k^4+\Delta_s^2+\Delta_2^2+d_0^2 \pm 2\kappa_0}, 
\eea
where
\bea
\nonumber \kappa_0=\sqrt{d_0(d_0k^4+2d_2\Delta_s\Delta_2)+\Delta_2^2(k^4+\Delta_0^2-d_2^2)}.\\
\eea
$|E_{-}(\boldsymbol{k})|$ vanishes when Eq.\ref{eq:rpf1} and Eq.\ref{eq:rpf2} are both satisfied. And ${\boldsymbol k}$-points, which satisfy both equalities, form gapless nodal rings. For the superconducting phase with ($d_{(3z^2-r^2,xy)}\!+\!id_{x^2-y^2}\!+\!s$) pairing, one can also investigate the gap structure by calculating $P(\boldsymbol{k})$. In this case, $k$-points, which make $P(\boldsymbol{k})=0$, form gapless surfaces.

\section{Gapless Bogoliubov Fermi surfaces and Topological Invariants} 
\label{sec:ti}

Recently Atland-Zirnbauer (AZ) classification has been extended for centrosymmetric system with inversion symmetry, $\mathcal{I}$ : so called AZ+$\mathcal{I}$ classfication.\cite{bzduvsek2017robust} Within AZ+$\mathcal{I}$ symmetry classes, our BdG Hamiltonian with ($d_{3z^2-r^2}\!+\!s$) pairing is classified as class DIII which support topologically charged nodal lines in spatial dimension $d=3$. Following Ref.\onlinecite{bzduvsek2017robust}, topological charge of nodal line can be characterized as $c^{d=3}_{\text{DIII}} \in \pi_1(M_{\text{DIII}}) =2\mathbb{Z}$ where $\pi_n(M_{CL})$ denotes n-th homotopy group of classifying space $M_{CL}$. To calculate the topological invariant of the nodal ring, we look into the BdG Hamiltonian with ($d_{3z^2-r^2}\!+\!s$) pairing. In the presence of time-reversal symmetry and particle-hole symmetry, the BdG Hamiltonian can be generally transformed with real manifolds of $\Delta$, as following.\cite{qi2010topological,yu2018singlet}
\begin{gather}
	\mathcal{H}(\boldsymbol{k})=\left(
	\begin{array}{cc}
		h_0(\boldsymbol{k}) & \Delta \\
		\Delta^\dagger & -h_0^T(\boldsymbol{k}) \\
	\end{array}
	\right)
	=\frac{1}{2}\left(
	\begin{array}{cc}
		1 & 1 \\
		i\gamma_{45} & -i\gamma_{45} \\
	\end{array}
	\right)
	\left(
	\begin{array}{cc}
		0 & h_0(\boldsymbol{k})-i\Delta\gamma_{45}^{\dagger} \\
		h_0(\boldsymbol{k})+i\Delta\gamma_{45}^{\dagger} & 0 \\
	\end{array}
	\right)
	\left(
	\begin{array}{cc}
		1 & -i\gamma_{45} \\
		1 & i\gamma_{45} \\
	\end{array}
	\right)
	\label{Eq:offd}.
\end{gather}
The off-diagonal component, $h_0(\boldsymbol{k})-i\Delta\gamma_{45}^{\dagger}$, can be explicitly expressed as follows,
\begin{gather}
	h_0(\boldsymbol{k})-i\Delta\gamma_{45}^{\dagger}=(c_0k^2+\sum_{i=1}^5 d_a(\boldsymbol{k})\gamma_a-\mu)-i(\gamma_{45}\Delta_s+\sum_a\gamma_a\gamma_{45}\Delta_a)\gamma_{45}^{\dagger}
	\\
	=(c_0k^2-\mu-i\Delta_s)+\sum_{i=1}^5 (d_a(\boldsymbol{k})-i\Delta_a)\gamma_a,
\end{gather}
and the corresponding eigenvalues are given as,
\begin{gather}
	\lambda_{\pm}=(c_0k^2-\mu-i\Delta_s)\pm\sqrt{\sum_{i=1}^5 (d_a(\boldsymbol{k})-i\Delta_a)^2}.
\end{gather}
For the superconducting phase with pure ($d_{3z^2-r^2}$) pairing, the order parameters are explicitly given as, $\Delta_s=0, \Delta=(0,\Delta_2,0,0,0)$. Then in the weak pairing limit, the eigenvalues, $\lambda_{\pm}$, can be expanded as a function of $\Delta_2$ up to the first order.
\begin{gather}
	\lambda_{\pm}\approx c_0k^2-\mu\pm(k^2-\frac{i(2k_z^2-k_x^2-k_y^2)\Delta_2}{2k^2}).
	\label{Eq:lambdaexpand}
\end{gather}
The first three terms are exactly the normal energy of the Luttinger Hamiltonian, which vanishes near the Fermi surface. i.e. it changes the sign at the Fermi surface. The third term is purely imaginary, which is proportional to the nodal structure, $2k_z^2-k_x^2-k_y^2$. As a result, the complex phase of $\lambda_{+(-)}$ winds by $2\pi$ as we encircle the nodal ring of the superconductor when $\mu>0(\mu<0)$. 
The winding of $\lambda$ manifests as the non-trivial topology of the nodal ring. To explicitly see this, we now define the topological invariant\cite{bzduvsek2017robust} along a loop that encircles the nodal ring as following,
\begin{gather}
	w=\frac{i}{2\pi}\int d\boldsymbol{k}\cdot \tr(q^\dagger(\boldsymbol{k}) \nabla_{\boldsymbol{k}} q(\boldsymbol{k})).
\end{gather} 
Here, $q(\boldsymbol{k})$ is given as,
\begin{gather}
	q(\boldsymbol{k})=\sum_n \frac{\lambda_n}{|\lambda_n|} | n\rangle \langle n| \equiv \sum_n e^{i\Phi_n}| n\rangle \langle n|
\end{gather}
where $| n\rangle$ is the eigenvectors of $h_0(\boldsymbol{k})-i\Delta\gamma_{45}^{\dagger}$. Without loss of generality, we can integrate on the $k_y=0$ plane where $k_x=k\sin\theta,k_z=k\cos\theta$.
\bea
\nonumber
w&=&\frac{i}{2\pi}\int_0^{2\pi} d\theta\cdot \tr(q^\dagger(\boldsymbol{k}) \partial_\theta q(\boldsymbol{k}))
=\sum_{m,n}\frac{i}{2\pi}\int_0^{2\pi} d\theta \tr(e^{-i\Phi_m}| m\rangle \langle m| \partial_\theta e^{i\Phi_n}| n\rangle \langle n|)
\\
&=&-\frac{1}{2\pi}\sum_n\int^{2\pi}_0 \partial_\theta \Phi_n+\sum_n \frac{i}{2\pi}\int_0^{2\pi} d\theta \tr(\langle n|\partial_\theta |n \rangle+(\partial_\theta\langle n|)|n\rangle )=-\frac{1}{2\pi}\sum_n\int^{2\pi}_0 \partial_\theta \Phi_n
=\pm 2,
\eea
where the signs for the top nodal ring ($k_z>0$) and the bottom nodal ring $(k_z<0)$ are opposite to each other. This winding number is topologically robust in the presence of the time-reversal and inversion symmetries. However, each nodal ring can shrink to a point at north pole and south pole (${\boldsymbol k} = \pm |k| \hat{z} $) when the $s$-wave pairing order parameter  $\Delta_s$ adiabatically deforms the nodal structure. The critical value of $\Delta_s$ where nodal rings shrink to nodal points can be derived by inspecting $P(k_x,k_y=0,k_z)$. The value is shown to be $\Delta_s=-\Delta_2$ when $c_0=0$ and $\Delta_2,\mu>0$. If we further increase $\Delta_s$, nodal points become gapped out on its own where each nodal point doesn't carry any topological charge at all.
Meanwhile, the nodal rings can also pair-annihilate at the equator ($k_z=0$) when $\Delta_s$ increases. The critical value of $\Delta_s$ that pair-annihilate nodal rings is given by $\Delta_s=\Delta_2\sqrt{\frac{\mu^2+\Delta_2^2}{4\mu^2+\Delta_2^2}}$ when $c_0=0$ and $\Delta_2,\mu>0$. 

\begin{figure}[h]
	\centering\includegraphics[width=0.7\textwidth]{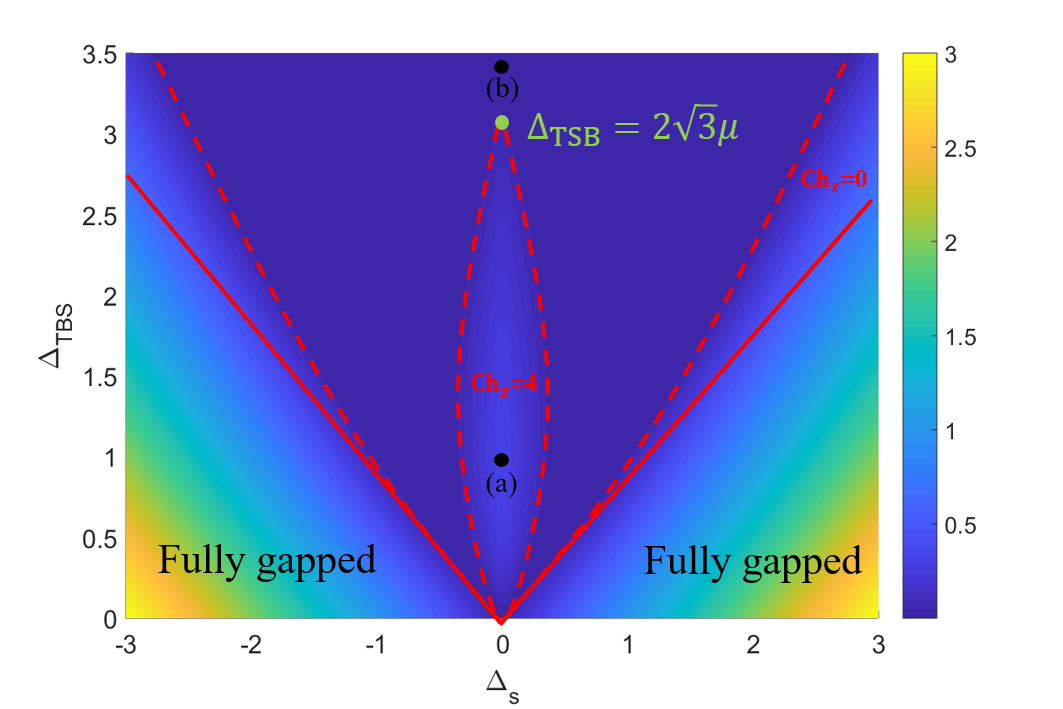}
	\caption{(Color Online) Phase diagram of superconducting gap structures and their topological properties as functions of $s$-wave pairing amplitude $\Delta_s$ and the time-reversal breaking $d$-wave component $\Delta_{TSB}$, when $\Delta_2=0$ and $\mu=1$. Color represents minimum value of $|E_n(k_x,k_y,k_z=0)|$ over $k_z=0$ plane among any $n$-th eigenvalue of $\mathcal{H}(\boldsymbol{k})$, $E_n({\boldsymbol{k}})$. $\text{Ch}_z$ denotes Chern number which is integrated over $k_z=0$ plane. Two black points represent special parameters that shows the gap structure as in Fig.\ref{fig:ch44}. Green point indicates the point where two Fermi pockets start to merge, thus it shows pure blue color for $\Delta_{TBS}>2\sqrt{3} \mu$ in the color map. Solid lines are for the phase boundaries between fully gapped and gapless superconducting phases and dashed lines are guides for the eyes for color map. 
	}
	\label{fig:tpd1}
\end{figure}

\begin{figure}[h]
	\subfloat[Gap structure of $\mathcal{H}(\boldsymbol{k})$ where $\Delta_{TSB}<2\sqrt{3}\mu$]{\label{fig:ch4}\includegraphics[scale=0.4]{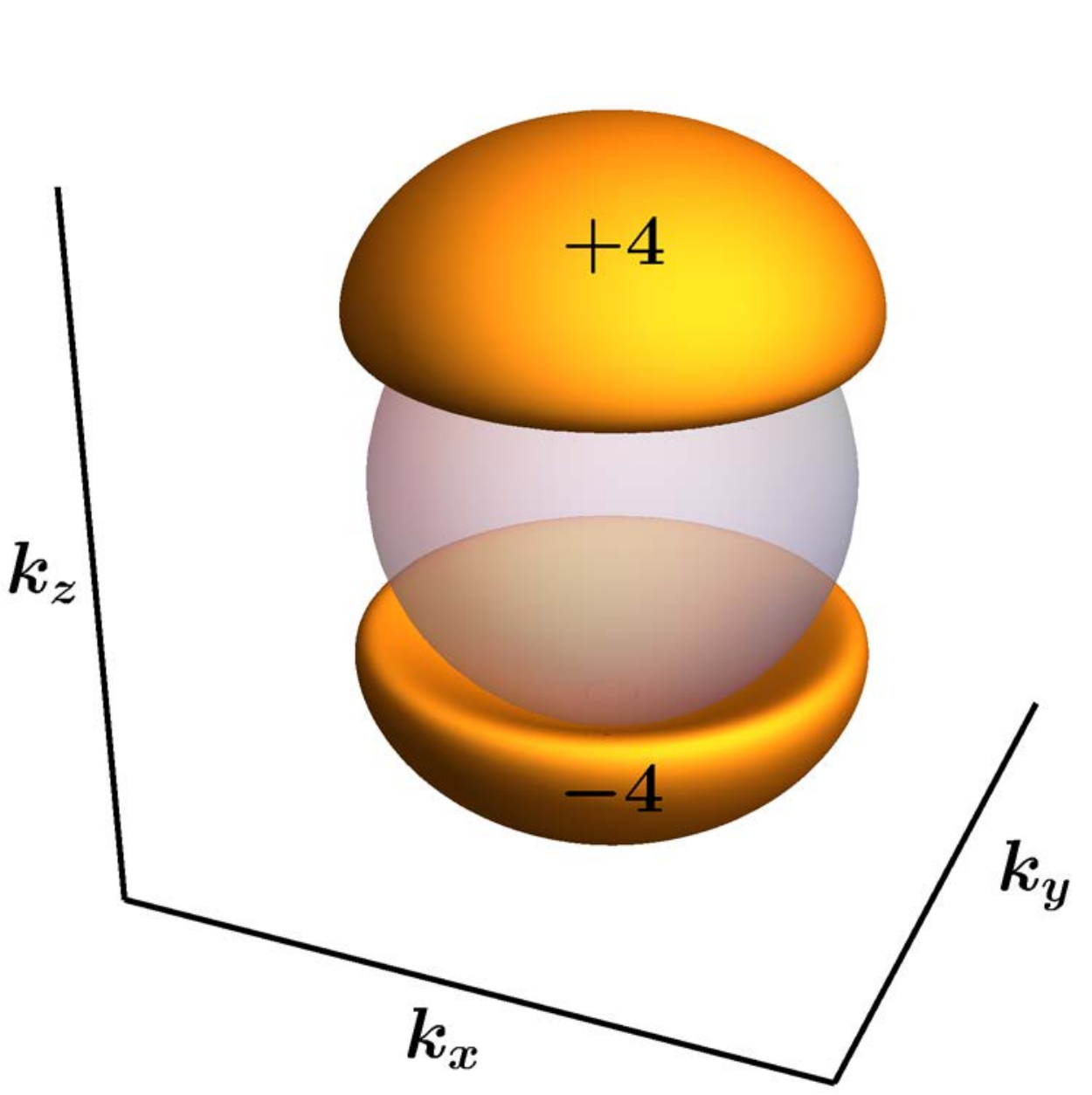}}~~~
	\subfloat[Gap structure of $\mathcal{H}(\boldsymbol{k})$ where $\Delta_{TSB}>2\sqrt{3}\mu$]{\label{fig:ch4-4}\includegraphics[scale=0.4]{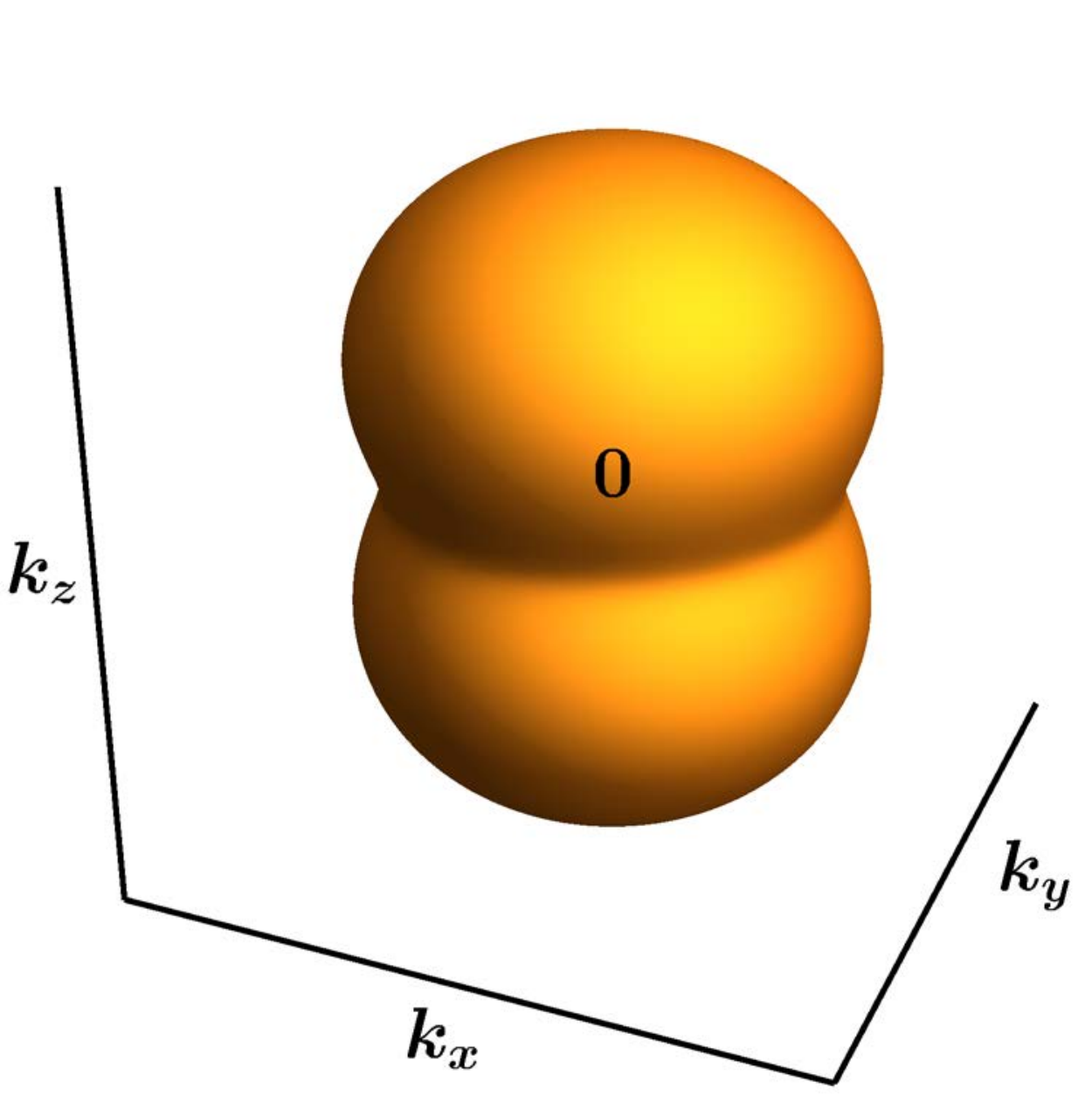}}~~~
	\caption{(Color Online) Evolution of Bogoliubov Fermi surface at (a) and (b) marked in Fig.\ref{fig:tpd1}. 
		(a) Fermi pocket located at $k_z>0~(k_z<0)$ has Chern number +4~(-4).
		(b) Fermi pocket has Chern number 0.}
	\label{fig:ch44}
\end{figure}

For the superconducting phase with ($d_{(3z^2-r^2,xy)}\!+\!id_{x^2-y^2}\!+\!s$) pairing, our BdG Hamiltonian is classified as class D which supports doubly charged surface nodes. Topological charges of the surface node, a Bogoliubov Fermi pocket, can be characterized as $c^{d=3}_{D} \in \pi_0(M_D) \bigoplus \pi_2(M_D)=\mathbb{Z}_2 \bigoplus 2\mathbb{Z}$. The 0-th homotopy charge $l$, is identified as $(-1)^l=\text{sgn}[P(\boldsymbol{k}_-)P(\boldsymbol{k}_+)]$ where $\boldsymbol{k}_-(\boldsymbol{k}_+)$ denotes $\boldsymbol{k}$ inside(outside) a Bogoliubov Fermi pocket. Meanwhile, the 2nd homotopy charge is related to the Chern number for a closed surface $S^2$ which enclose a nodal surface. For numerical calculation of Chern number, we adopt the Kubo type formula given below.
\bea
\nonumber \text{Ch}=\frac{i}{2\pi}\sum_{E_n<E_F<E_m}\oint\limits_{S^2}d^2\boldsymbol{s}(\boldsymbol{k}) \cdot \frac{\bra{u_n(\boldsymbol{k})}\nabla_{\boldsymbol{k}}\mathcal{H}(\boldsymbol{k})\ket{u_m(\boldsymbol{k})}\times\bra{u_m(\boldsymbol{k})}\nabla_{\boldsymbol{k}}\mathcal{H}(\boldsymbol{k})\ket{u_n(\boldsymbol{k})}}{(E_n(\boldsymbol{k})-E_m(\boldsymbol{k}))^2},
\eea
where $\mathcal{H}(\boldsymbol{k})$ indicates the BdG Hamiltonian matrix at a given ${\boldsymbol k}$, $E_F$ is the Fermi level and $\ket{u_n(\boldsymbol{k})}$ and $E_n({\boldsymbol k})$ denote the $n$th Bloch state and energy respectively. To investigate the topological properties of the ($d_{(3z^2-r^2,xy)}+id_{x^2-y^2}+s$) pairing, we decompose the $d$-wave order parameter as below.
\bea
\vec{\Delta}=\Delta_{2}(0,1,0,0,0)+\Delta_{TSB}(i, 0, 0, 0, 1)
\eea
Instead of calculating the Chern number of each Bogoliubov Fermi pocket, we integrate out the Berry curvature over $k_z=0$ plane and define the Chern number Ch$_z$. On the $k_z=0$ plane, the BdG Hamiltonian can be simplified and block-diagonalized as the following two sub-matrices with $c_0=0$.
\bea
h_{BdG1}&=&
\left(
\begin{array}{cccc}
	\frac{1}{2} \left(-k_x^2-k_y^2-2 \mu \right) & \frac{1}{2} \sqrt{3} (k_x-i k_y)^2 & 0 & i \left(\Delta _s+\Delta _{2}\right) \\
	\frac{1}{2} \sqrt{3} (k_x+i k_y)^2 & \frac{1}{2} \left(k_x^2+k_y^2-2 \mu \right) & i \left(\Delta _{2}-\Delta _s\right) & 2 \Delta _{TSB} \\
	0 & i \left(\Delta _s-\Delta _{2}\right) & -\frac{k_x^2}{2}+\mu -\frac{k_y^2}{2} & \frac{1}{2} \sqrt{3} (k_x+i k_y)^2 \\
	-i \left(\Delta _s+\Delta _{2}\right) & 2 \Delta _{TSB} & \frac{1}{2} \sqrt{3} (k_x-i k_y)^2 & \frac{1}{2} \left(k_x^2+k_y^2\right)+\mu  \\
\end{array}
\right),
\\
h_{BdG2}&=&
\left(
\begin{array}{cccc}
	\frac{1}{2} \left(k_x^2+k_y^2-2 \mu \right) & -\frac{1}{2} \sqrt{3} (k_x-i k_y)^2 & 0 & i \left(\Delta _s-\Delta _{2}\right) \\
	-\frac{1}{2} \sqrt{3} (k_x+i k_y)^2 & \frac{1}{2} \left(-k_x^2-k_y^2-2 \mu \right) & -i \left(\Delta _s+\Delta _{2}\right) & -2 \Delta_{TSB} \\
	0 & i \left(\Delta _s+\Delta _{2}\right) & \frac{1}{2} \left(k_x^2+k_y^2\right)+\mu  & -\frac{1}{2} \sqrt{3} (k_x+i k_y)^2 \\
	i \left(\Delta _{2}-\Delta _s\right) & -2 \Delta _{TSB} & -\frac{1}{2} \sqrt{3} (k_x-i k_y)^2 & -\frac{k_x^2}{2}+\mu -\frac{k_y^2}{2} \\
\end{array}
\right).
\eea
If a small positive value of $\Delta_{TSB}$ is turned on while $\Delta_s=\Delta_{2}=0$, the Chern number of Fermi pocket, which is located at $k_z>0~ (k_z<0)$, becomes $+4~(-4)$ as shown in Fig.\ref{fig:ch4}. When $\Delta_{TSB} \geq 2\sqrt{3}\mu$ with $\Delta_s,\Delta_{2}=0$, two Fermi
pockets with opposite Chern numbers merge forming a single Fermi pocket with Chern number 0 as shown in Fig.\ref{fig:ch4-4}. Fig.\ref{fig:tpd1} represents the phase diagram of superconducting gap structures and thier topological properties as functions of $\Delta_s$ and $\Delta_{TSB}$ when $\Delta_2=0$ and $\mu=1$. Here, two black points indicate specific parameters that show gap structure as shown in Fig.\ref{fig:ch44}. Color indicates the minimum value of $|E_n(k_x,k_y,k_z=0)|$ over $k_z=0$ plane among any $n$-th eigenvalue of $\mathcal{H}(\boldsymbol{k})$, $E_n({\boldsymbol{k}})$, which indicates the evolution of Bogoliubov Fermi surface topology.
Solid lines indicate the phase boundaries between the fully gapped superconductor and the superconductor with Bogoliubov Fermi pockets. Dashed lines are guides for the eyes for color map, where Bogoliubov Fermi pockets touch $k_z=0$ plane or not. 

\begin{figure}[h]
	\centering\includegraphics[width=0.7\textwidth]{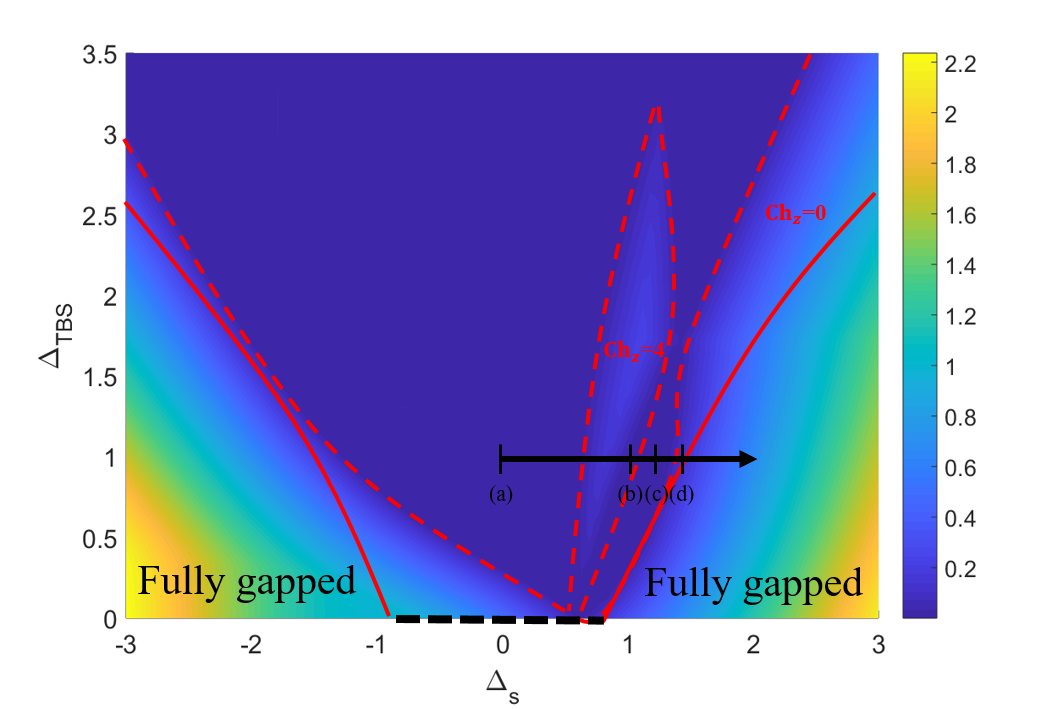}
	\caption{(Color Online) Phase diagram of superconducting gap structures and their topological properties as functions of $\Delta_s$ and $\Delta_{TSB}$ when $\Delta_2=\mu=1$. Color represents minimum value of $|E_n(k_x,k_y,k_z=0)|$ over $k_z=0$ plane among any $n$-th eigenvalue of $\mathcal{H}(\boldsymbol{k})$, $E_n({\boldsymbol{k}})$. $\text{Ch}_z$ denotes Chern number which is integrated over $k_z=0$ plane. Red solid lines are for the phase boundaries between fully gapped and gapless superconducting phases and red dashed lines are guides for the eyes for color map. Black dashed line at $\Delta_{TBS}=0$ is for the superconducting phase where nodal rings are stabilized. Black arrow indicates the parameter line where topological phase transition can occur. Evolution of the gap structure is shown in Fig.\ref{fig:ch22}.}
	\label{fig:tpd2}
\end{figure}

\begin{figure}[h]
	\subfloat[Gap structure of $\mathcal{H}(\boldsymbol{k})$ where $\Delta_s=0$]{\label{fig:ch00}\includegraphics[scale=0.3]{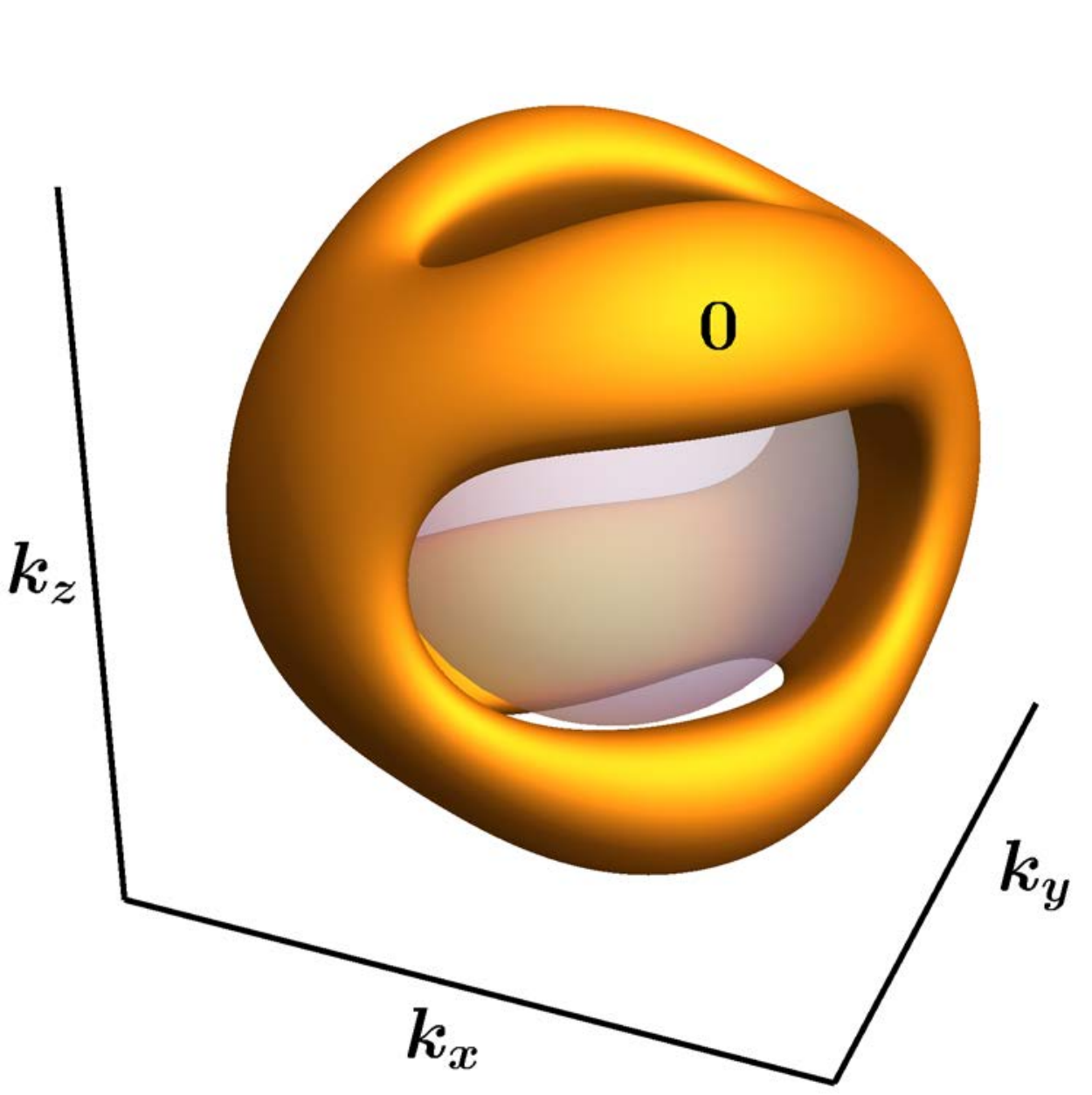}}~~~
	\subfloat[Gap structure of $\mathcal{H}(\boldsymbol{k})$ where $\Delta_s=1$]{\label{fig:ch2}\includegraphics[scale=0.3]{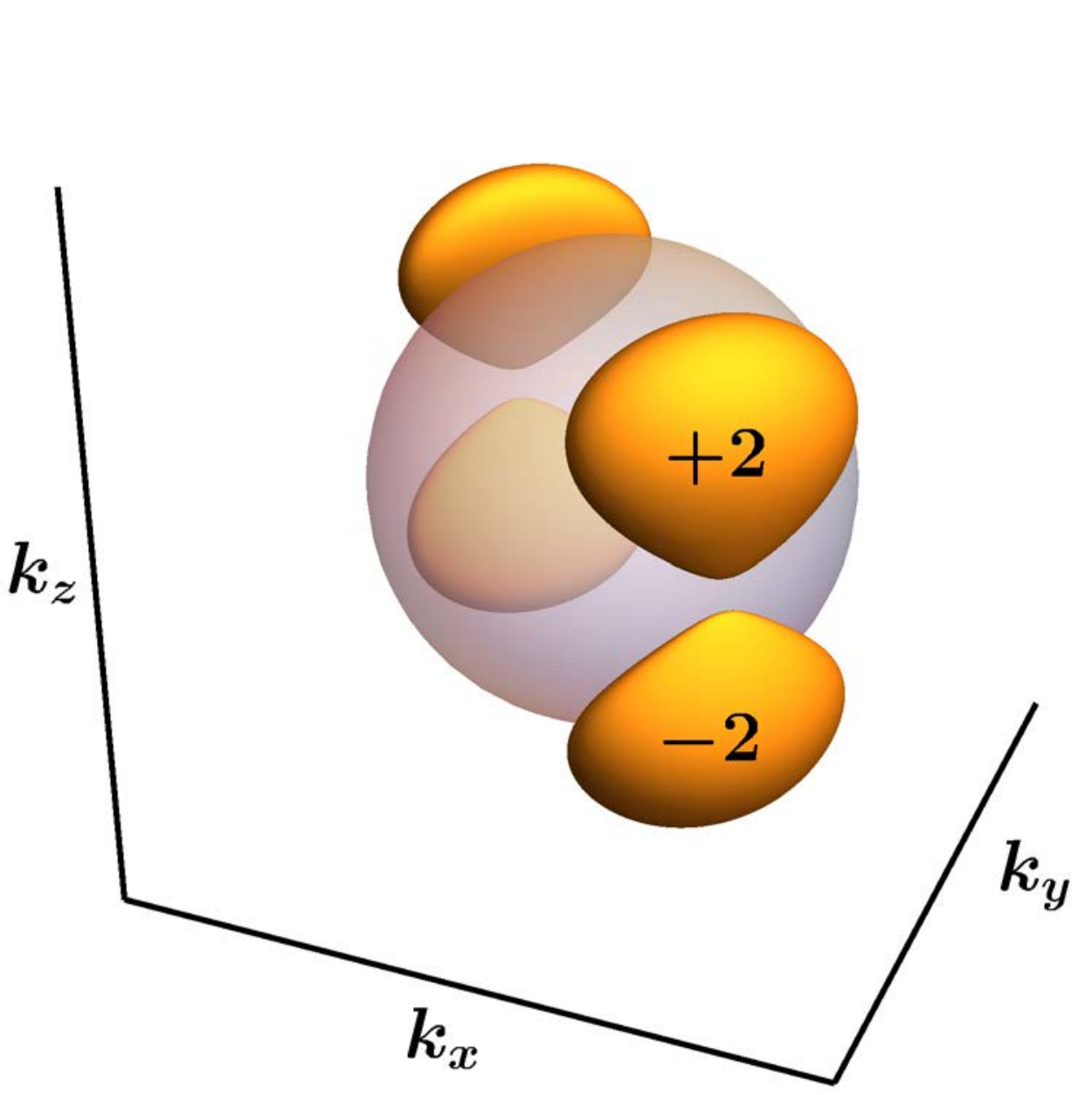}}~~~
	\subfloat[Gap structure of $\mathcal{H}(\boldsymbol{k})$ where $\Delta_s=1.2$]{\label{fig:ch2-2}\includegraphics[scale=0.3]{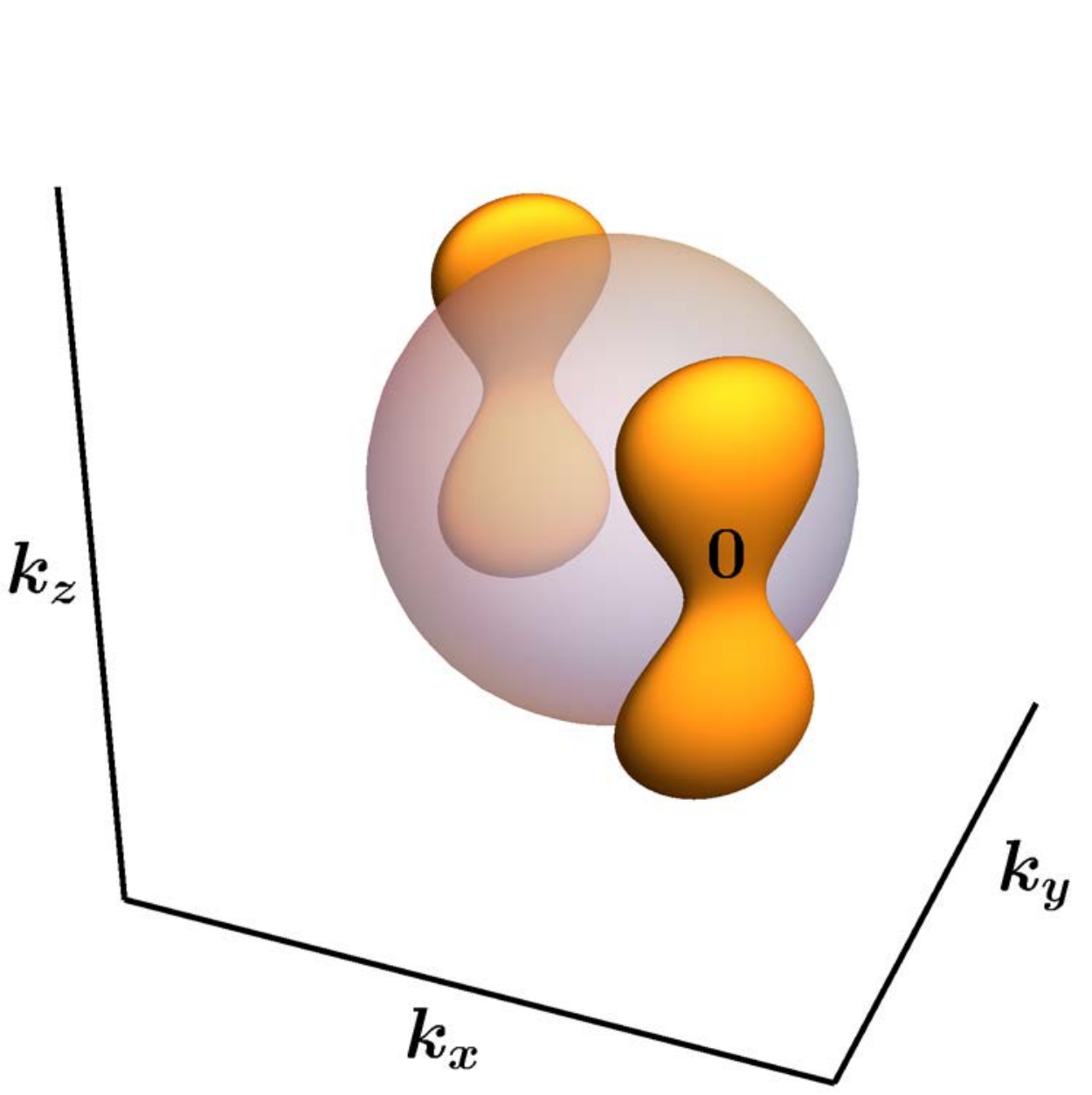}}~~~
	\subfloat[Gap structure of $\mathcal{H}(\boldsymbol{k})$ where $\Delta_s=1.4$]{\label{fig:ch0}\includegraphics[scale=0.3]{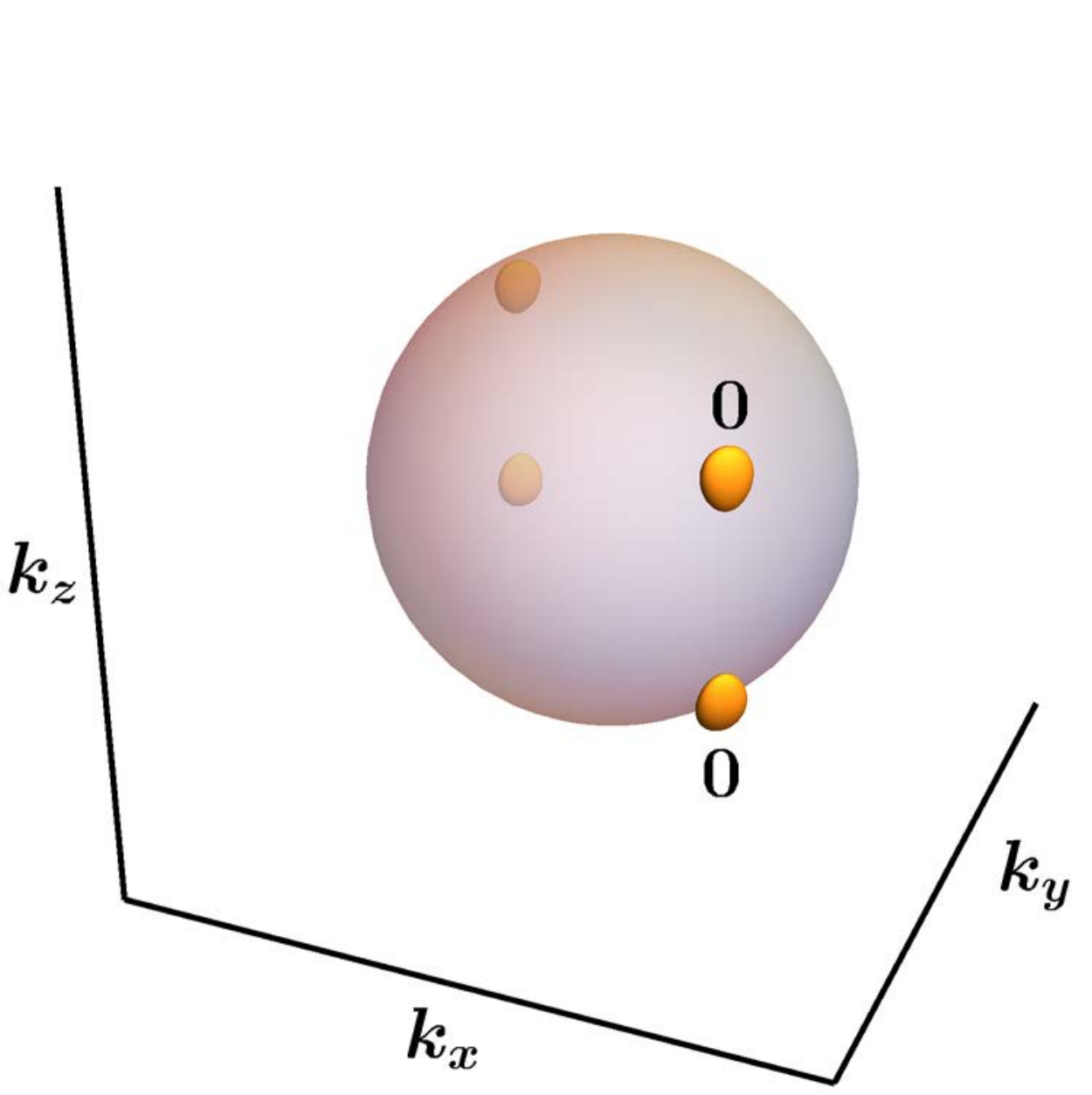}}~~~
	\caption{(Color Online) Evolution of Bogoliubov Fermi surface along black arrow shown in Fig.\ref{fig:tpd2}. 
		(a) Fermi surface has Chern number 0.
		(b) Fermi pockets located at $k_z>0~(k_z<0)$ have Chern number +2~(-2). 
		(c) Merge of two Fermi pockets having Chern numbers 0. 
		(d) Each Fermi pocket has Chern number 0. }
	\label{fig:ch22}
\end{figure}

Similarly, Fig.\ref{fig:tpd2} shows the phase diagram for non-zero $\Delta_2$, set $\Delta_2=\mu=1$ as functions of $\Delta_s$ and $\Delta_{TSB}$. We can see the topological phase transitions when $\Delta_{TSB}=1$ by tuning $0\leq\Delta_s\leq2$ as black arrow in Fig.\ref{fig:tpd2}. 
First, we can see the Bogoliubov Fermi surface with $\text{Ch}=0$ when $\Delta_s=0$ as in Fig.\ref{fig:ch00}. When $\Delta_s=1$, there exist two pairs of Fermi Pockets, which are located at $k_z>0(k_z<0)$ with $\text{Ch}=+2(-2)$ as shown in Fig.\ref{fig:ch2}. By increasing $\Delta_s$, we can see that each pair of Fermi pockets with opposite Chern numbers merge forming a single Fermi pocket with $\text{Ch}=0$ as in Fig.\ref{fig:ch2-2}. Further increment of $\Delta_s$ lead each pocket to split again into two Fermi pockets with $\text{Ch}=0$ as shown in Fig.\ref{fig:ch0}. Eventually, the system becomes fully-gapped when $\Delta_s$ is large enough. We also check that nodal ring persist when $-\Delta_2<\Delta_s<\Delta_2\sqrt{\frac{\mu^2+\Delta_2^2}{4\mu^2+\Delta_2^2}}$ and $\Delta_{TSB}=0$ which is shown as a black dashed line in Fig.\ref{fig:tpd2}. 

\vskip 0.2cm

\end{document}